\documentstyle{article}

\def\noheaderplainsetup{

\topmargin=0pt \headheight=0pt \headsep=0pt  \oddsidemargin=0pt \evensidemargin=0pt  \textheight=8.94truein \textwidth=6.5truein}   

\noheaderplainsetup

\begin{document}

\newcommand{\ar}{\mbox{\bf CLA1}}
\newcommand{\artwo}{\mbox{\bf CLA2}}
\newcommand{\arthree}{\mbox{\bf CLA3}} 
\newcommand{\cltw}{\mbox{\bf CL12}} 

\newcommand{\elz}[1]{\mbox{$\parallel\hspace{-3pt} #1 \hspace{-3pt}\parallel$}} 
\newcommand{\elzi}[1]{\mbox{\scriptsize $\parallel\hspace{-3pt} #1 \hspace{-3pt}\parallel$}}
\newcommand{\gj}[1]{{\bf #1}} 
\newcommand{\emptyrun}{\langle\rangle} 
\newcommand{\oo}{\bot}            
\newcommand{\pp}{\top}            
\newcommand{\xx}{\wp}               
\newcommand{\legal}[2]{\mbox{\bf Lr}^{#1}_{#2}} 
\newcommand{\win}[2]{\mbox{\bf Wn}^{#1}_{#2}} 
\newcommand{\seq}[1]{\langle #1 \rangle}           

\newcommand{\gneg}{\mbox{\small $\neg$}}                  
\newcommand{\intimpl}{\mbox{\hspace{2pt}$\circ$\hspace{-0.14cm} \raisebox{-0.058cm}{\Large --}\hspace{2pt}}}
\newcommand{\pintimpl}{\mbox{\hspace{2pt}\raisebox{0.033cm}{\tiny $>$}\hspace{-0.18cm} \raisebox{-0.043cm}{\large --}\hspace{2pt}}}
\newcommand{\mli}{\hspace{2pt}\mbox{\small $\rightarrow$}\hspace{2pt}}                      
\newcommand{\cla}{\mbox{$\forall$}}      
\newcommand{\cle}{\mbox{$\exists$}}        
\newcommand{\mld}{\hspace{2pt}\mbox{\small $\vee$}\hspace{2pt}}     
\newcommand{\mlc}{\hspace{2pt}\mbox{\small $\wedge$}\hspace{2pt}}   
\newcommand{\mlci}{\hspace{2pt}\mbox{\footnotesize $\wedge$}\hspace{2pt}}   
\newcommand{\ade}{\mbox{\large $\sqcup$}}      
\newcommand{\ada}{\mbox{\large $\sqcap$}}      
\newcommand{\add}{\hspace{2pt}\mbox{\small $\sqcup$}\hspace{2pt}}                     
\newcommand{\adc}{\hspace{2pt}\mbox{\small $\sqcap$}\hspace{2pt}} 
\newcommand{\adci}{\hspace{2pt}\mbox{\footnotesize $\sqcap$}\hspace{2pt}}              
\newcommand{\clai}{\forall}     
\newcommand{\clei}{\exists}        
\newcommand{\tlg}{\bot}               
\newcommand{\twg}{\top}               
\newcommand{\fintimpl}{\mbox{\hspace{2pt}$\bullet$\hspace{-0.14cm} \raisebox{-0.058cm}{\Large --}\hspace{-6pt}\raisebox{0.008cm}{\scriptsize $\wr$}\hspace{-1pt}\raisebox{0.008cm}{\scriptsize $\wr$}\hspace{4pt}}}
\newcommand{\fintimpli}{\mbox{\hspace{1pt}{\scriptsize $\bullet$}\hspace{-0.14cm} \raisebox{-0.048cm}{--}\hspace{-5pt}\raisebox{0.008cm}{\tiny $\wr$}\hspace{-1pt}\raisebox{0.008cm}{\tiny $\wr$}\hspace{3pt}}}
\newcommand{\sfbr}{\mbox{\hspace{2pt}$\bullet$\hspace{-0.14cm} \raisebox{-0.058cm}{\Large --}\hspace{-6pt}\hspace{1pt}\raisebox{0.008cm}{\scriptsize $\wr$}\hspace{4pt}}}
\newcommand{\fbr}{\mbox{\hspace{2pt}$\bullet$\hspace{-0.14cm} \raisebox{-0.058cm}{\Large --}\hspace{2pt}}}
\newcommand{\sfbri}{\mbox{\mbox{\hspace{1pt}{\scriptsize $\bullet$}\hspace{-0.14cm} \raisebox{-0.048cm}{--}\hspace{-4pt}\hspace{-1pt}\raisebox{0.008cm}{\tiny $\wr$}\hspace{3pt}}}}
\newcommand{\fbri}{\mbox{\mbox{\hspace{1pt}{\scriptsize $\bullet$}\hspace{-0.14cm} \raisebox{-0.048cm}{--}\hspace{-4pt}\hspace{-1pt}\hspace{3pt}}}}
\newcommand{\col}[1]{\mbox{$#1$:}}


\newtheorem{theoremm}{Theorem}[section]
\newtheorem{factt}[theoremm]{Fact}
\newtheorem{definitionn}[theoremm]{Definition}
\newtheorem{lemmaa}[theoremm]{Lemma}
\newtheorem{propositionn}[theoremm]{Proposition}
\newtheorem{conventionn}[theoremm]{Convention}
\newtheorem{examplee}[theoremm]{Example}
\newtheorem{exercisee}[theoremm]{Exercise}

\newenvironment{definition}{\begin{definitionn} \em}{ \end{definitionn}}
\newenvironment{theorem}{\begin{theoremm}}{\end{theoremm}}
\newenvironment{lemma}{\begin{lemmaa}}{\end{lemmaa}}
\newenvironment{fact}{\begin{factt}}{\end{factt}}
\newenvironment{proposition}{\begin{propositionn} }{\end{propositionn}}
\newenvironment{convention}{\begin{conventionn} \em}{\end{conventionn}}
\newenvironment{proof}{ {\bf Proof.} }{\  $\Box$ \vspace{.1in} }
\newenvironment{example}{\begin{examplee} \em}{\end{examplee}}
\newenvironment{exercise}{\begin{exercisee} \em}{\end{exercisee}}

\title{Towards  applied theories based on computability logic}
\author{Giorgi Japaridze}  
\date{}
\maketitle
\begin{abstract} {\em Computability logic} (CL) is a recently launched program for redeveloping logic as a formal theory of computability, as opposed to the formal theory of truth that logic has more traditionally been. Formulas in it represent computational problems,  ``truth'' means existence of an algorithmic solution, and proofs encode such solutions. Within the line of research devoted to finding axiomatizations for ever more expressive fragments of CL, the present paper introduces a new deductive system {\bf CL12} and proves its soundness and completeness with respect to the semantics of CL. Conservatively extending classical predicate calculus and offering considerable additional expressive and deductive power, {\bf CL12} presents a reasonable, computationally meaningful, constructive alternative to classical logic as a basis for applied theories. To obtain a model example of such theories, this paper rebuilds the traditional, classical-logic-based  Peano arithmetic into a computability-logic-based counterpart.  Among the purposes of the present contribution is to provide a starting point for what, as the author wishes to hope, might become a new line of research with a potential of interesting  findings --- an exploration of the presumably quite unusual metatheory of CL-based arithmetic and other CL-based applied systems.

\end{abstract}


\


\section{Introduction}\label{intr}

{\em Computability logic} (CL) is a semantical platform and program for redeveloping logic as a formal theory of computability, as opposed to a formal theory of truth which logic has more traditionally been.  It sees formulas as interactive computational problems viewed as games played by a machine ($\twg$) against the environment ($\tlg$), where logical operators stand for operations on games and ``truth'' is understood as existence of an algorithmic winning strategy. 
Numerous papers  \cite{Jap03}-\cite{Japfour}  have been published on this subject in recent years, and the present reader is assumed to be familiar with the basic philosophy, motivations and techniques of CL. Otherwise, he or she is recommended  to take a look at the first 10 sections of \cite{Japfin} for an introduction and  survey. 
 
So far all technical efforts within the CL project have been focused on developing CL as a pure logic, typically through  constructing  axiomatizations for various fragments of it. The contribution presented in the first part of this paper is in the same style. It introduces a new reduction operation $\fintimpl$, similar to the earlier known $\intimpl$ but stronger in that the antecedent is allowed to be replicated only a finite number of times.  With $\fintimpl$ syntactically treated as a separator of the two parts of a sequent, a  sequent-calculus-style deductive system \cltw\ is then constructed, and a proof of its soundness and completeness with respect to the semantics of CL is provided. All atoms of the language of \cltw\ are  elementary, and its logical vocabulary, other than $\fintimpl$, contains the earlier known $\gneg,\mlc,\mld,\adc,\add,\cla,\cle,\ada,\ade$.

It should however be remembered that the ultimate purpose of logic is providing an intellectual tool for navigating the real life. As such, first and foremost it should be able to successfully serve as a basis for  applied (substantial) theories. 
 The possibility and expediency of basing applied theories on CL has been repeatedly pointed out in earlier papers (\cite{Jap03,Japtcs,Japic,Japfin}), but no concrete attempts had been undertaken  to do so until now. The second part of the present article makes long overdue (by the standards of the young but rapidly evolving CL) initial steps in this direction --- the direction which appears to be full of promise, mystery and the excitement of upturning virgin soil. For the first time in the  history of the project, it constructs a computability-logic-based applied theory, specifically, Peano arithmetic based on \cltw\ instead of classical logic. This system, named \ar, is meant to be a model example of a theory of this kind, and a starting point for future explorations that, as the author has reasons to believe, may result in some interesting findings. 

Generally, the nonlogical axioms  of a computability-logic-based applied  theory would be any collection of (formulas expressing) problems whose algorithmic solutions are known. Sometimes, together with nonlogical axioms, we may also have nonlogical rules of inference, preserving the property of computability. An example of such a rule is the {\em constructive induction rule} of \ar.  Then, the soundness of the corresponding underlying axiomatization of CL (such as \cltw\ in the present case) --- which usually comes in the strong form called {\em uniform-constructive soundness} --- guarantees that every theorem $T$ of the theory also has an algorithmic solution and that, furthermore, such
a solution can be effectively constructed from a proof of $T$. Does not this look like exactly what the constructivists have been calling for?! 

On the applied side, the above fact makes CL-based theories problem-solving tools: finding a solution for a given problem reduces to expressing the problem in the language of the theory, and finding a proof of that problem. An algorithmic solution for the problem then automatically comes together with such a proof.

\section{The dfb-reduction operation, informally}\label{ss2}

As noted, the reader is assumed to be familiar with the basics of CL.  Any unknown or forgotten terms and symbols used here can be looked up in \cite{Japfin}, which comes with a convenient glossary.\footnote{The glossary for the published version of \cite{Japfin} is given at the end of the {\em book} (rather than {\em article}), on pages 371-376. The reader may instead use the preprint version of \cite{Japfin}, available at http://arxiv.org/abs/cs.LO/0507045 The latter includes both the main text and the glossary.} 
 
 Consider the problem $\ada x\ade  y(y=x^2)$. Anyone who knows the definition of $x^2$ (but perhaps does not know the definition of $x\times y$, or is unable to compute the multiplication function for whatever reason) would be able to solve the following problem, as it is about reducing the consequent to the antecedent: 
\begin{equation}\label{april15}
\ada z\ada u\ade v(v= z\times u)\ \mli \ \ada x\ade y(y=x^2).
\end{equation}
A machine's winning strategy for the above goes like this. Wait till the environment specifies a value $n$ for $x$, i.e. asks ``what is the square of $n$?''. Do not try to immediately answer this question, but rather specify the same value $n$ for both $z$ and $u$, thus asking the counterquestion ``what is $n$ times $n$?''. The environment will have to provide a correct answer $m$  to this counterquestion, i.e., specify $v$ as $m$ where $m=n\times n$, or else it loses. Then, specify $y$ as $m$, and rest your case. Note that, in this solution, the machine did not have to compute multiplication, doing which had become environment's responsibility. The machine only correctly reduced the problem of computing square to the problem of computing product, which made it the winner.

Even though  (\ref{april15}) expresses a ``very easily solvable'' problem, this formula is still not logically valid. The success of the reduction strategy of the consequent to the antecedent that we provided for it relies on the nonlogical fact that $x^2=x\times x$. That strategy would fail in a general case where the meanings of $x^2$ and $x\times x$ may not necessarily be the same. On the other hand, the goal of CL as a general-purpose problem-solving tool should be to allow us find purely logical solutions, i.e., solutions that do not require any special, domain-specific knowledge and (thus) would be good no matter what the particular predicate or function symbols of the formulas mean. Any knowledge that might be relevant should be explicitly stated and included either in the antecedent of a given formula or in the set of axioms (``implicit antecedents'' for every potential formula) of a CL-based theory. 
In our present case,  formula (\ref{april15}) easily turns into a logically valid one by adding, to its antecedent,  the definition of square in terms of multiplication:
\begin{equation}\label{april16}
\cla w (w^2=w\times w) \mlc \ada z\ada u\ade v(v= z\times u)\ \mli \ \ada x\ade y(y=x^2).
\end{equation}
The strategy that we provided  for (\ref{april15}) is just as good for (\ref{april16}), with the difference that it is successful for (\ref{april16}) no matter what $x^2$ and $z\times u$ mean, whereas, in the case of (\ref{april15}), it was guaranteed to be successful only under the standard arithmetic interpretations of the square and product functions.

Let us now look at the following modification of  (\ref{april16}):
\begin{equation}\label{april16a}
\cla w \bigl(w^3=(w\times w)\times w\bigr) \mlc \ada z\ada u\ade v(v= z\times u)\ \mli \ \ada x\ade y(y=x^3).
\end{equation}
Here is a pseudo-strategy for (\ref{april16a}) in the style of the (real) strategy that we gave for (\ref{april15}) and (\ref{april16}). Wait till the environment specifies a value $n$ for $x$. Then, using the resource $\ada z\ada u\ade v(v= z\times u)$ given in the antecedent, make the environment compute $m$ with $m=n\times n$. Then, again using the same resource, make the environment further compute $m\times n$, and use the resulting value to specify $y$ in the consequent. 

Of course, there was some cheating here. Namely, after the first usage of  $\ada z\ada u\ade v(v= z\times u)$, it will have evolved to the elementary game $m=n\times n$ and hence will be no longer available as a product-computing resource.  The above pseudo-strategy works not for (\ref{april16a}) but for the following, weaker problem, which has not one but rather two copies of the product-computing resource in the antecedent: 
\[\cla w \bigl(w^3=(w\times w)\times w\bigr) \mlc \ada z\ada u\ade v(v= z\times u)\mlc \ada z\ada u\ade v(v= z\times u)\ \mli \ \ada x\ade y(y=x^3).\]
In this particular case we got lucky in finding an easy way around,  as computing cube only requires computing product a fixed number of times, which can be accounted for by including, in the antecedent, the appropriate number of copies of the relevant resource. But what if the problem was not about computing cube or square, but rather computing $x^h$ for a ($\ada$-bound) variable $h$? 

Here we see  a need for considering  a reduction operation that, unlike $\mli$, permits using the antecedental resources any finite number of times. Several 
earlier-studied weak reduction operations, including $\intimpl$ and $\pintimpl$, do allow repeated usage of the antecedent. But, imposing no restrictions on the quantity of repetitions and thus permitting infinitely many reusages, for most purposes they  turn out to be weaker than necessary. At the same time, they tend to be very hard to tame syntactically/axiomatically.  The operation $\fintimpl$ that we introduce in this paper and call 
{\bf dfb-reduction} appears to be semantically optimal in this respect, and also syntactically better behaved. 

It is applied to two components: $\vec{A}$, called the {\bf antecedent}, and $B$, called the {\bf succedent}. The result is written $\vec{A}\fintimpl B$. Here $B$ is any game, while $\vec{A}$ is any finite sequence of games. More precisely, $\vec{A}$ is not really a sequence but rather a binary tree with games at its leaves, as will be seen later from the strict definitions. But, in the present informal explanation, we can safely ignore this technical detail and identify such a tree with the sequence of the games at its leaves.  
 
Intuitively, the meaning of 
\begin{equation}\label{uu1}A_1,\ldots,A_n\fintimpl B\end{equation}
is similar to that of 
\begin{equation}\label{uu2}A_1\mlc\ldots\mlc A_n\mli B,\end{equation}
that is, the task of the machine is to win in $B$ as long as the environment, playing in the role of $\pp$, wins against the machine (the latter playing in the  role of $\oo$) in each $A_i$. This amounts to solving problem $B$ while having the problems $A_1,\ldots,A_n$ as computational resources. The difference between (\ref{uu1}) and (\ref{uu2}) is that, in the former, the machine has the additional capability to replicate, at any time, any of the ``conjuncts'' of the antecedent (in the form to which it has evolved  by that time rather than in the original form), and insert the new copy of it as a new ``conjunct'' next to the old copy.  A restriction here is that the machine is allowed to do such replications only a finite  number of times, or else it will be considered to have lost the game. 

As an aside, ``dfb'' stands for {\em double-finite branching}. One can define two other, similar but weaker versions of reduction: {\em single-finite branching} (sfb) reduction $\sfbr$ and (simply) {\em finite branching} (fb) reduction $\fbr$. What makes $\sfbr$ weaker than $\fintimpl$ is that, in the former, 
winning in the succedent automatically makes $\pp$ the winner, even if it did infinitely many replications in the antecedent. 
 And what makes $\fbr$ further weaker than $\sfbr$ is that, in the former, it does not matter at all whether the machine has made finitely or infinitely many replications; however, when determining the winner, only the copies of the antecedental resources that emerged as a result of finitely many replications will be looked at. We will be exclusively concerned with dfb-reduction in this paper, so sfb- and fb-reductions can be safely forgotten.

Expressions of the form $\vec{A}\fintimpl B$ we call {\bf sequents}. If $\pp$ has an algorithmic winning strategy for the game represented by such an expression, we say that $B$ is {\bf dfb-reducible} to $\vec{A}$, and call such a strategy (just as the problem $\vec{A}\fintimpl B$ itself) a {\bf dfb-reduction} of $B$ to $\vec{A}$.  The following sequent represents the problem of dfb-reducing the problem of computing cube to the problem of computing multiplication and the definition of cube: 
 
\begin{equation}\label{april16b}
\cla w \bigl(w^3=(w\times w)\times w\bigr), \ \ada z\ada u\ade v(v= z\times u)\ \fintimpl \ \ada x\ade y(y=x^3).
\end{equation}
Here is a winning strategy for (\ref{april16b}), which succeeds for any possible meanings of $x^3$ and $z\times u$, so not only does (\ref{april16b}) have an algorithmic solution, but, in fact, it is a logically valid sequent. The strategy is to first wait till the environment picks a value $n$ for $x$. This brings the game down to 
\[\cla w \bigl(w^3=(w\times w)\times w\bigr),\  \ada z\ada u\ade v(v= z\times u)\ \fintimpl \ \ade y(y=n^3).\]
Now the machine replicates the second resource of the antecedent, bringing the game down to 
\[\cla w \bigl(w^3=(w\times w)\times w\bigr), \ \ada z\ada u\ade v(v= z\times u), \ \ada z\ada u\ade v(v= z\times u)\ \fintimpl \ \ade y(y=n^3).\]
After this, the machine specifies both $z$ and $u$ as $n$ in the second resource of the antecedent, further bringing the game down to 
\[\cla w \bigl(w^3=(w\times w)\times w\bigr),\  \ade v(v= n\times n),\  \ada z\ada u\ade v(v= z\times u)\ \fintimpl \ \ade y(y=n^3).\]
To avoid a loss, the environment will have to respond by choosing a (correct) value $m$ for $v$ in the same component, and the game evolves to 
\[\cla w \bigl(w^3=(w\times w)\times w\bigr), \ m= n\times n, \ \ada z\ada u\ade v(v= z\times u)\ \fintimpl \ \ade y(y=n^3).\]
Now the machine specifies $z$ and $u$ as $m$ and $n$ in the third resource of the antecedent, bringing the game down to 
\[\cla w \bigl(w^3=(w\times w)\times w\bigr),  \ m= n\times n, \ \ade v(v= m\times n)\ \fintimpl \ \ade y(y=n^3).\]
Again, the environment will have to respond by selecting a value $k$ for $v$, and we get
\[\cla w \bigl(w^3=(w\times w)\times w\bigr), \ m= n\times n,\  k= m\times n\ \fintimpl \ \ade y(y=n^3).\]
Finally, the machine specifies $y$ as $k$ and, having brought the play down to the following true elementary game, celebrates victory:
\[\cla w \bigl(w^3=(w\times w)\times w\bigr), \ m= n\times n, \ k= m\times n\ \fintimpl \ k=n^3.\]
Note that the succedent of the above sequent is a logical consequence (in the classical sense) of the formulas of the antecedent. This can be seen to imply that, as promised, the success of our strategy in no way depends on the meanings of $x^3$ and $z\times u$, so that the strategy is, in fact, a ``purely logical'' solution. Again, among the purposes of computability logic is to serve as a tool for finding such ``purely logical'' solutions, so that it can be applied to any domain of study rather than specific domains such as that of arithmetic, and to arbitrary meanings of nonlogical symbols rather than particular meanings such as that of the multiplication function for the symbol $\times$. 
 
Remember the concept of Turing reducibility. A function $f(x)$ is said to be {\em Turing reducible} to functions $g_1(x),\ldots,g_n(x)$ iff there is a Turing machine that computes $f(x)$ with oracles for $g_1(x),\ldots,g_n(x)$. This is easily seen to mean nothing but the dfb-reducibility of $\ada x\ade y\bigl(y=f(x)\bigr)$ to 
the sequence \[\ada x\ade y\bigl(y=g_1(x)\bigr),\ \ldots,\ \ada x\ade y\bigl(y=g_n(x)\bigr).\] Thus, dfb-reducibility is a conservative generalization of Turing reducibility. The former, of course, is more general, as it is not limited to problems of the form $\ada x\ade y\bigl(y=h(x)\bigr)$ as Turing reducibility is.

\section{Dfb-reduction and a few other concepts defined formally}\label{cg}

Remember from \cite{Japfin} that a {\em constant game} is a pair $A=(\legal{A}{},\win{A}{})$, where $\legal{A}{}$ is the set of legal runs of $A$ and $\win{A}{}$ is a function telling us which player ($\pp$ or $\oo$) has won a given run. Further remember that the letter $\xx$ is always used as a variable for players, and $\overline{\xx}$ (sometimes written as $\gneg \xx$) means $\xx$'s adversary. 

We also remind the reader the definition of the operation of prefixation: 

\begin{definition}\label{prfx}
Let $A$ be a constant game and $\Phi$ a legal position of $A$. The game 
$\seq{\Phi}A$ is defined by: $\legal{\seq{\Phi}A}{}=\{\Gamma\ |\ \seq{\Phi,\Gamma}\in\legal{A}{}\}$;
$\win{\seq{\Phi}A}{}\seq{\Gamma}=\win{A}{}\seq{\Phi,\Gamma}$.
\end{definition}

We say that a constant game $A$ is {\bf finite-depth} iff there is an integer $d$ such that no legal run of $A$ contains more than $d$ (lab)moves. The smallest of such integers $d$ is called the {\bf depth} of $A$. An {\bf elementary game} is a game of depth $0$.

Our main focus in this paper will be on finite-depth games. This restriction of focus makes many definitions and proofs simpler. Namely, in order to define a game operation $O(A_1,\ldots,A_n)$ applied to such games,\footnote{The same, in fact, holds for the wider class of games termed in \cite{Jap03} {\em perifinite-depth}. These are games with finite yet perhaps arbitrarily long  legal runs.}  it suffices  to specify the following:

\begin{description}
\item[(i)] Who wins $O(A_1,\ldots,A_n)$ if no moves are made, i.e., the value of $\win{O(A_1,\ldots,A_n)}{}\emptyrun$.
\item[(ii)] What are the {\bf initial legal (lab)moves}, i.e., the elements of  $\{\xx\alpha\ |\ \seq{\xx\alpha}\in\legal{O(A_1,\ldots,A_n)}{}\}$, and to 
what game  is $O(A_1,\ldots,A_n)$ brought down after such an initial legal labmove $\xx\alpha$ is made. Recall that, by saying that a given labmove $\xx\alpha$ brings a given game $A$ down to $B$, we mean that $\seq{\xx\alpha}A=B$.  
\end{description}
Then, the set of legal runs of $O(A_1,\ldots,A_n)$ will be uniquely defined, and so will be the winner in every run of the game. If, however, infinite legal runs may also occur, then the following additional stipulation will  be required:  

\begin{description}
\item[(iii)] How to determine the value of $\win{O(A_1,\ldots,A_n)}{}\seq{\Gamma}$ when $\Gamma$ is an infinite legal run of $O(A_1,\ldots,A_n)$.
\end{description}

For instance, a few basic game operations of CL already known from the earlier literature, can be (re)defined as follows: 

\begin{definition}\label{op} Let $A$, $B$, $A_0,A_1,\ldots$ be finite-depth constant games, and $n\in\{1,2,\ldots\}$.\vspace{9pt}

\noindent 1. $\gneg A$ is defined by: 
\begin{quote}\begin{description}
\item[(i)] $\win{\gneg A}{}\emptyrun=\xx$ iff $\win{A}{}\emptyrun=\overline{\xx}$. 
\item[(ii)] $\seq{\xx\alpha}\in\legal{\gneg A}{}$ iff $\seq{\overline{\xx}\alpha}\in\legal{A}{}$. Such an initial legal labmove $\xx\alpha$ brings the game down to 
$\gneg \seq{\overline{\xx}\alpha}A$.\vspace{5pt}
\end{description}\end{quote}

\noindent 2. $A_0\adc\ldots\adc  A_n$ is defined by: 
\begin{quote}\begin{description}
\item[(i)] $\win{A_0\adci\ldots\adci  A_n}{}\emptyrun=\pp$. 
\item[(ii)] $\seq{\xx\alpha}\in\legal{A_0\adci\ldots\adci  A_n}{}$ iff $\xx=\oo$ and $\alpha=i\in\{0,\ldots,n\}$.  Such an initial legal labmove $\oo i$ brings the game down to 
$A_i$.\vspace{5pt} 
\end{description}\end{quote}

\noindent 3. $A_0\mlc\ldots\mlc A_n$ is defined by: 
\begin{quote}\begin{description}
\item[(i)] $\win{A_0\mlci\ldots\mlci  A_n}{}\emptyrun=\pp$ iff, for each $i\in\{0,\ldots,n\}$,  $\win{A_i}{}\emptyrun=\pp$. 
\item[(ii)] $\seq{\xx\alpha}\in\legal{A_0\mlci\ldots\mlci  A_n}{}$ iff $\alpha=i.\beta$, where $i\in\{0,\ldots,n\}$ and $\seq{\xx\beta}\in\legal{A_i}{}$. Such an initial legal labmove $\xx i.\beta$ brings the game down to  
\( A_0\mlc\ldots\mlc A_{i-1}\mlc \seq{\xx\beta}A_i\mlc A_{i+1}\mlc\ldots\mlc A_n.\vspace{3pt}\) 
\end{description}\end{quote}

\noindent 4. $A_0\add\ldots\add A_n$ and $A_0\mld\ldots\mld  A_n$
are defined exactly as $A_0\adc\ldots\adc A_n$ and $A_0\mlc\ldots\mlc  A_n$, respectively, only with ``$\pp$" and ``$\oo$" interchanged.\vspace{7pt}

\noindent 5. The infinite $\adc$-conjunction $A_0\adc A_1\adc\ldots$ is defined exactly as $A_0\adc\ldots\adc A_n$, only with ``$i\in\{0,1,\ldots\}$" instead of ``$i\in\{0,\ldots,n\}$". Similarly for the infinite version of $\add$.\vspace{7pt}


\end{definition}

To define the (new, never studied before) operation $\fintimpl$ of {\bf dfb-reduction}, we need some preliminaries.  What we call a  {\bf tree of games} is a structure defined inductively as an element of the smallest set satisfying the following conditions:
\begin{itemize}
\item Every game $A$ is a tree of games. The one-element sequence $\seq{A}$ is said to be the {\bf yield} of such a tree, and the {\bf address} of $A$ in this tree is the empty bit string. 
\item Whenever $\cal A$ is a tree of games with yield $\seq{A_1,\ldots,A_m}$ and $\cal B$ is a  tree of games with yield $\seq{B_1,\ldots,B_n}$, the pair ${\cal A}\circ{\cal B}$ is a tree of games with yield $\seq{A_1,\ldots,A_m,B_1,\ldots,B_n}$. The address of each $A_i$  in this tree is $0w$, where $w$ is the address of $A_i$ in $\cal A$. Similarly,  the address of each $B_i$ is $1w$, where $w$ is the address of $B_i$ in $\cal B$.
\end{itemize}

Example: Where $A,B,C,D$ are games, $(A\circ B)\circ(C\circ(A\circ D))$ is a tree of games with yield $\seq{A,B,C,A,D}$. The address of the first $A$ of the yield, to which we may as well refer as the first {\bf leaf} of the tree, is $00$, the address of the second leaf $B$ is $01$, the address of the third leaf $C$ is $10$, the address of the fourth leaf $A$ is $110$, and the address of the fifth leaf $D$ is $111$.

Note that $\circ$ is not an operation on games, but just a symbol used instead of the more common comma to separate two parts of a pair. And a tree of games itself is not a game, but a collection of games arranged into a certain structure, just as a sequence of games is not a game but a collection of games arranged as a list. 

For bit strings $u$ and $w$, we will be writing $u\preceq w$ to indicate that $u$ is a (not necessarily proper) {\bf prefix} (initial segment) of $w$.

\begin{definition}\label{op1} 
Let  $A_1,\ldots,A_n,B$ be  finite-depth constant games, and $\cal T$ be a tree of games with yield $\seq{A_1,\ldots,A_n}$ ($n\geq 1$). Let $w_1,\ldots,w_n$ be the addresses of $A_1,\ldots,A_n$ in $\cal T$, respectively. The game   ${\cal T}\fintimpl B$ is defined by:
\begin{description}
\item[(i)] $\win{{\cal T}\fintimpli B}{}\emptyrun=\pp$ iff $\win{B}{}\emptyrun=\pp$ or, for some $i\in\{1,\ldots,n\}$, $\win{A_i}{}\emptyrun=\oo$. In other words, 
$\win{{\cal T}\fintimpli B}{}\emptyrun=\win{A_1\wedge\ldots\wedge A_n\rightarrow B}{}\emptyrun$.
\item[(ii)] $\seq{\xx\alpha}\in\legal{{\cal T}\fintimpli B}{}$ iff one of the following conditions is satisfied: 
\begin{enumerate}
\item $\xx\alpha=\xx S.\beta$ (here ``$S$'' stands for ``succedent''), where $\seq{\xx\beta}\in\legal{B}{}$. Such a labmove $\xx S.\beta$ brings the game down to  \({\cal T}\fintimpl \seq{\xx\beta}B.\)
\item $\xx\alpha=\xx u.\beta$, where $u\preceq w_i$ for at least one $i\in\{1,\ldots,n\}$ and, for each  $i$ with $u\preceq w_i$,  $\seq{\overline{\xx}\beta}\in\legal{A_i}{}$. Such a labmove $\xx u.\beta$ brings the game down to  \({\cal T}'\fintimpl B\), where ${\cal T}'$ is the result of replacing $A_i$ by $\seq{\overline{\xx}\beta}A_i$ in $\cal T$ for each $i$ with $u\preceq w_i$. 
\item $\xx\alpha=\pp \col{w_i}$, where $i\in\{1,\ldots,n\}$. We call such a move a {\bf replicative (lab)move}. It brings the game down to  ${\cal T}'\fintimpl B$, where ${\cal T}'$ is the result of replacing $A_i$ by $(A_i\circ A_i)$ in $\cal T$.
\end{enumerate}
\item[(iii)] If $\Gamma$ is a legal run of ${\cal T}\fintimpl B$ with infinitely many replicative moves (which is the only case when $\Gamma$ can and will be infinite), then $\win{{\cal T}\fintimpli B}{}\seq{\Gamma}=\oo$.
\end{description}
\end{definition}

\begin{example} \label{ex37}
Let $p,q,r,s$ be any propositions (elementary constant games). Let $G$ be the following game:
\[(p\adc q)\adc (r\adc s)\fintimpl p\mlc \bigl((q\adc r)\mlc (q\adc s)\bigr).\]
And let $\Gamma$ be the run 
\[\seq{\pp \col{},\ \oo S.1.1.1,\ \pp 0.1,\ \pp 0.1,\ \oo S.1.0.0,\ \pp \col{1}, \ \pp 1.0,\ \pp 10.0,\ \pp 11.1}.\]

\noindent $\Gamma$ is a legal run of $G$. Below we trace, step by step, the effects of its moves on $G$. 

The 1st (lab)move $\pp \col{}$, meaning replicating the (only) leaf of the antecedental tree, with the address of that leaf being the empty bit string, 
brings $G$ down to --- in the sense that $\seq{\pp\col{}}G$ is --- the following game:
 \[\bigl((p\adc q)\adc (r\adc s)\bigr)\circ\bigl((p\adc q)\adc (r\adc s)\bigr) \fintimpl p\mlc \bigl((q\adc r)\mlc (q\adc s)\bigr).\]

The 2nd move $\oo S.1.1.1$ of $\Gamma$, meaning making the move $\oo 1.1.1$ in the succedent, meaning making the move $\oo 1.1$ in the right $\mlc$-conjunct of the succedent, meaning making the move $\oo 1$ in the right $\mlc$-conjunct of the right $\mlc$-conjunct of the succedent, meaning selecting the right $\adc$-conjunct there,  further brings the game down to --- in the sense that $\seq{\pp\col{},\oo S.1.1.1}G$ is --- the following game:
  \[\bigl((p\adc q)\adc (r\adc s)\bigr)\circ\bigl((p\adc q)\adc (r\adc s)\bigr) \fintimpl p\mlc \bigl((q\adc r)\mlc s\bigr).\]

The 3rd move $\pp 0.1$ of $\Gamma$, meaning making the move $1$ in the $0$-addressed leaf of the antecedental tree, meaning selecting the right $\adc$-conjunct there,  further brings the game down to --- i.e., $\seq{\pp\col{},\oo S.1.1.1,\pp 0.1}G$ is --- the following game:
 \[(r\adc s)\circ\bigl((p\adc q)\adc (r\adc s)\bigr) \fintimpl p\mlc \bigl((q\adc r)\mlc s\bigr).\]

The 4th move $\pp 0.1$ further brings the game down to 
 \[s\circ\bigl((p\adc q)\adc (r\adc s)\bigr) \fintimpl p\mlc \bigl((q\adc r)\mlc s\bigr).\]

The 5th move $\oo S.1.0.0$ further brings the game down to 
 \[s\circ\bigl((p\adc q)\adc (r\adc s)\bigr) \fintimpl p\mlc (q\mlc s).\]

The 6th move $\pp \col{1}$, replicating the $1$-addressed leaf of the antecedental tree, further brings the game down to 
 \[s\circ\Bigl(\bigl((p\adc q)\adc (r\adc s)\bigr)\circ \bigl((p\adc q)\adc (r\adc s)\bigr)\Bigr) \fintimpl p\mlc (q\mlc s).\]

The 7th move $\pp 1.0$,  whose effect is the same as the effect of the two consecutive moves $\pp 10.0$ and $\pp 11.0$ would be, further brings the game down to 
 \[s\circ\bigl((p\adc q)\circ (p\adc q)\bigr) \fintimpl p\mlc (q\mlc s).\]

The 8th move $\pp 10.0$ further brings the game down to 
 \[s\circ\bigl(p\circ (p\adc q)\bigr) \fintimpl p\mlc (q\mlc s).\]

The last, 9th move $\pp 11.1$ further brings the game down to the eventual position/game $\seq{\Gamma}G$, which is
 \[s\circ (p\circ q) \fintimpl p\mlc (q\mlc s).\]

Note that this play is won by $\pp$ no matter what particular propositions $p,q,r$ and $s$ are. If, however, the run had stopped after the 8th move, it could have been lost by $\pp$, namely, $\pp$ would lose in case $s$ and $p$ are true but $q$ is false.
\end{example}

For the rest of this paper we fix two infinite sets of expressions: the set 
$\{v_0,v_1,\ldots\}$ of {\bf variables} and the set $\{0,1,\ldots\}$ (decimal numerals) of {\bf constants}. Without loss of generality here we assume that the above collection of constants is exactly the universe of discourse
--- i.e., the set over which the variables range --- in all cases that we consider. As always, by a {\bf valuation} we mean 
a function $e$ that sends each variable $x$ to a constant $e(x)$. In these terms, a classical predicate $p$ can be understood as 
a function that sends each valuation $e$ to a proposition, i.e. constant predicate.   Similarly, what we call a (simply) game sends valuations to constant games. Here we reproduce a formal definition of this concept and associated notation from \cite{Japfin}: 

\begin{definition}\label{ngame}
A {\bf game} is a function $A$ from valuations to constant games. We write $e[A]$ (rather than $A(e)$) to denote the constant game returned by $A$ for valuation $e$. Such a constant game $e[A]$ is said to be an {\bf instance} of $A$. 
For readability, we usually write $\legal{A}{e}$ and $\win{A}{e}$ instead of $\legal{e[A]}{}$ and $\win{e[A]}{}$.
\end{definition}

Where $n$ is a natural number, we say that a game $A$ is {\bf $n$-ary} iff there is are $n$ variables such that, for any two valuations $e_1$ and $e_2$ that agree on all those variables, we have $e_1[A]=e_2[A]$. Generally, a game that is $n$-ary for some $n$, is said to be {\bf finitary}.

We say that a game $A$ {\bf depends} on a variable $x$ iff there are two valuations  $e_1,e_2$ that agree on all variables except $x$ such that $e_1[A]\not=e_2[A]$. An $n$-ary game thus depends on at most $n$ variables. And constant games are nothing but $0$-ary games.

Just as the Boolean operations straightforwardly extend from propositions to all predicates, our operations 
$\gneg,\mlc,\mld,\mli,\adc,\add,\fintimpl$ extend from constant games to all games. This is done by simply stipulating that $e[\ldots]$ commutes with all of those operations, as well as with $\circ$: $\gneg A$ is 
the game such that, for every $e$, $e[\gneg A]=\gneg e[A]$; $A\adc B$ is the game such that,
for every $e$, $e[A\adc B]=e[A]\adc e[B]$; ${\cal T}\fintimpl B$ is the game such that,
for every $e$, $e[{\cal T}\fintimpl B]=e[{\cal T}]\fintimpl e[B]$, where $e[{\cal T}]$ is the result of replacing in $\cal T$ every game $A$ of the yield by $e[A]$; etc. The operation of prefixation also extends to nonconstant games: whenever $\Phi$ is a legal position of every instance of a game $A$,  $\seq{\Phi}A$ should be understood as the unique game such that, for every $e$, $e[\seq{\Phi}A]=\seq{\Phi}e[A]$. 

We remind the reader that the choice quantifiers can be defined as follows: 

\[\ada xA(x)=_{def}A(0)\adc A(1)\adc A(2)\adc\ldots ;\]  \[\ade xA(x)=_{def}A(0)\add A(1)\add A(2)\add\ldots.\]  

A game $A$ is said to be {\bf unistructural} iff, for any two valuations $e_1$ and $e_2$, we have $\legal{A}{e_1}=\legal{A}{e_2}$. Of course, all constant or elementary games are unistructural. It can also be easily seen that all our game operations preserve the unistructural property of games. For the purposes of the present paper, considering only unistructural games would be sufficient. 

A unistructural game is said to be {\em finite-depth} iff so are all of its instances. We (re)define the remaining operations $\cla$ and $\cle$ only for unistructural, finite-depth games:

\begin{definition}\label{op5} Let $A(x)$ be a unistructural, finite-depth game.\vspace{9pt}

\noindent 1. $\cla x A(x)$ is defined by stipulating that, for every valuation $e$ and move $\alpha$, we have: 
\begin{quote}\begin{description}
\item[(i)] $\win{\clai x A(x)}{e}\emptyrun=\pp$ iff, for every constant $c$, $\win{A(c)}{e}\emptyrun=\pp$. 
\item[(ii)] $\seq{\xx\alpha}\in\legal{\clai x A(x)}{e}$ iff $\seq{\xx\alpha}\in\legal{A(x)}{e}$. Such an initial legal labmove $\xx\alpha$ brings the game $e[\cla x A(x)]$ down to 
$e[\cla x\seq{\xx\alpha}A(x)]$.\vspace{5pt}
\end{description}\end{quote}
\noindent 2. $\cle x A(x)$ is defined in exactly the same way, only with $\pp$ and $\oo$ interchanged.  
\end{definition}

\section{Block-move machines}\label{icp}

As always, we agree that the term ``{\bf computational problem}'', or simply ``{\bf problem}'', is a synonym of ``static game'' (see Section \ref{ss13} 
 for a definition). All meaningful and reasonable examples of games --- including all elementary games --- are static, and the class of static games is closed under all of our  game operations. This fact for $\fintimpl$ is proven in the appendix-style Section \ref{ss13} of the present paper, and for all other operations it has been proven in \cite{Jap03} (Theorem 14.1).

The paper \cite{Jap03} defines two models of interactive computation, called the {\em hard-play machine} ({\em HPM}) and the {\em easy-play machine} ({\em EPM}). We remind the reader that  both are sorts of Turing machines with the capability of making moves, and have three tapes: the ordinary read/write {\em work tape}, and the read-only {\em valuation} and {\em run} tapes. The 
valuation tape contains a full description of some valuation $e$, and its content remains fixed throughout the work of the machine. In this paper we only consider finitary games, and a simplifying assumption that we are making is that the valuation tape only lists the values of finitely many (relevant) variables, the rest of the tape being blank, and the value of $e$ at all unlisted variables assumed to be $0$. As for the run tape, it serves as a dynamic input, at any time spelling     
the current position. Every time one of the players makes a move, that move --- with the corresponding label --- is automatically appended to the content of this tape. 
In the HPM model, there is no restriction on the frequency of environment's moves.
In the EPM model, on the other hand, the machine has full control over the speed of its adversary: the environment can 
(but is not obligated to) make a (one single) move only when the machine explicitly allows it to do so --- the event that we call {\bf granting permission}. The only ``fairness" requirement that such a machine is expected to satisfy is that it should grant permission every once in a while; how long that ``while" lasts, however, is totally up to the machine.  
The HPM and EPM models seem to be two extremes, yet, according to Theorem 17.2 of \cite{Jap03}, they yield the same class of 
winnable static games. The present paper will only deal with the EPM model, so let us take a little closer look at it.

Let $\cal M$ be an EPM. A {\em configuration}\label{y10} of $\cal M$ is defined in the standard way: this is a full description of the (``current") state of the machine, the locations of its three scanning heads
and the contents of its three tapes.
The {\em initial configuration} on valuation $e$ is the configuration where $\cal M$ is in its start state, the valuation tape spells $e$, each scanning head is looking at the first cell of the corresponding tape, and the work and run tapes are empty. A configuration $C'$ is said to be a {\em successor} of configuration $C$ in $\cal M$ if $C'$ can legally follow $C$ in the standard --- standard for multitape Turing machines --- sense, based on the transition function (which we assume to be deterministic) of the machine and accounting for the possibility of nondeterministic updates 
--- depending on what move $\oo$ makes or whether it makes a move at all --- of the content of the run tape when $\cal M$ grants permission. Technically, granting permission happens by entering one of the specially designated states called ``permission states". And $\cal M$ makes a move $\alpha$ by constructing it at the beginning of its work tape and then entering one of the specially designated states called ``move states". A {\bf computation branch} of $\cal M$ on $e$ is a sequence of configurations of $\cal M$ where the first
configuration is the   initial configuration on $e$ and every other configuration is a successor of the previous one.
Thus, the set of all computation branches on $e$ captures all possible scenarios (on valuation $e$) corresponding to different behaviors by $\oo$. Such a branch is said to be {\bf fair} iff permission is granted infinitely many times in it.
Each computation branch $B$ of $\cal M$ incrementally spells --- in the obvious sense --- a run $\Gamma$ on the run tape, which we call the {\bf run spelled by $B$}. Then, for a game $A$, we write ${\cal M}\models_e A$ and say that  $\cal M$ {\bf wins $A$ on valuation $e$}     to mean that, whenever $\Gamma$  is the run spelled by some computation branch $B$ of $\cal M$ on $e$ and $\Gamma$ is not $\oo$-illegal,\footnote{Remember that a {\bf $\xx$-illegal} run is an illegal run where the last move of the shortest illigal initial segment is $\xx$-labeled.} then  branch $B$ is fair and $\win{A}{e}\seq{\Gamma}=\pp$. We write ${\cal M}\models A$ and say that $\cal M$ {\bf wins} ({\bf computes, solves}) $A$ iff ${\cal M}\models_e A$ for every valuation $e$. Finally, we write $\models A$ and say that 
$A$ is {\bf computable} iff there is an EPM $\cal M$ with ${\cal M}\models A$. 

The class of static games is robust with respect to choosing different models of computation or different modes of playing games. Specifically, procedural decisions regarding whether players are allowed to move at any time (as in the HPM model), or only when $\pp$ so decides (as in the EPM model), or only when $\oo$ so decides, or whether they should take alternating turns, do not affect the class of winnable static games. Further, if the players move in an alternating order, it turns out to be irrelevant which player starts the game, or how many moves at a time either player is allowed to make.  

In the present paper, together with the EPM model, we will also be using the new (never defined before) model of interactive computation that we call {\bf BMEPM} (``Block-Move EPM''). A BMEPM is the same as an EPM, with the only difference that here either player can make any finite number of moves at once. That is, the machine can make several moves on a given clock cycle (say, by first listing those moves on its work tape and then entering a move state); and the environment, too, can make any finite number of moves (including no moves at all) when granted permission. All concepts and notations that we use for EPMs straightforwardly extend to BMEPMs as well. The following proposition officially establishes the equivalence between EPMs and BMEPMs for static games:

\begin{proposition}\label{bmepm}
A static game $A$ is computable (i.e., there is an EPM that wins $A$) iff there is a BMEPM that wins $A$. Furthermore:

1.  There is an effective procedure that converts any BMEPM $\cal M$ into an EPM $\cal E$ such that, for any static game $A$ and any valuation $e$, if ${\cal M}\models_e A$, then ${\cal E}\models_e A$. 

2. And vice versa: there is an effective procedure that converts any EPM $\cal E$ into a BMEPM $\cal M$ such that, for any static game $A$ and any valuation $e$, if ${\cal E}\models_e A$, then ${\cal M}\models_e A$.

\end{proposition}

\begin{proof} {\em Clause 1}: Consider an arbitrary BMEPM $\cal M$. The corresponding EPM $\cal E$ is a machine that, with a valuation $e$ spelled on its valuation tape, works by simulating $\cal M$ with the same $e$ spelled on the imaginary valuation tape of the latter.  Every time this simulation shows that $\cal M$ grants permission, $\cal E$ also grants permission, and then feeds the environment's response (which can be a single move or no move) back to the simulated $\cal M$. And, every time the simulation shows that $\cal M$ makes a block $\alpha_1,\ldots,\alpha_n$ of moves, $\cal E$ makes the same $n$ moves --- only, one by one rather than the whole block at once --- in its real play. Obviously, the run $\Gamma$ generated by $\cal E$ in this play is a run that could have been generated by $\cal M$ (in precise terms, $\Gamma$ is the run spelled by some computation branch of $\cal M$ on $e$). So, if $\cal M$ wins a game $A$ on valuation $e$, so does $\cal E$.

  {\em Clause 2}: Consider an arbitrary EPM $\cal E$. The corresponding BMEPM $\cal M$ is a machine that, with  $e$ spelled on its valuation tape, works by simulating $\cal E$ with the same $e$ spelled on the valuation tape of the latter.  Every time this simulation shows that
$\cal E$ makes a move, $\cal M$ makes the same move. And every time the simulation shows that  $\cal E$ grants permission, so does $\cal M$. If, in response, the environment makes no move, or makes a single move, then the same response is fed back to the simulated $\cal E$. Suppose now the environment responds by a block $\alpha_1,\ldots,\alpha_n$ of moves ($n\geq 2$). In this case, $\cal M$ feeds back the response $\alpha_1$ to the simulated $\cal E$. It continues simulation (mimicking ${\cal E}$'s moves in its real play) until $\cal E$ grants permission again, in which case $\cal M$ (without granting permission in the real play) simply feeds the environment's earlier response $\alpha_2$ into the simulation, and so on until all of the $n$ moves are processed in the same manner, after which $\cal M$ resumes the ordinary simulation. It is not hard to see that the run $\Delta$ generated in this play by $\cal M$ is a $\pp$-delay of the run $\Gamma$ generated by the simulated $\cal E$.  
 So, if $\cal E$ wins a static game $A$ on valuation $e$, so does $\cal M$.
\end{proof}

\section{Formulas and sequents}\label{ss6}

The  first-order  language of (the fragment of) CL that we consider in this section has infinitely many  {\em variables} and  {\em constants} --- the ones fixed in Section \ref{cg}. Next, it has infinitely many 
$n$-ary {\em nonlogical predicate symbols} and $n$-ary  {\em function symbols} for each integer $n\geq 0$. Finally, it has one binary  {\em logical predicate symbol} $=$. {\em Terms} are defined in the usual way, and so are {\em formulas}, except that the connectives now are $\twg,\tlg$ ($0$-ary), $\gneg$ ($1$-ary),  
$\mlc$,$\mld$,$\adc$,$\add$ ($2$-ary), and the quantifiers are $\cla,\cle,\ada,\ade$. Also,  $\gneg$ is only allowed on atomic formulas. In any other case it should be understood as a standard abbreviation, and so should $\mli$. The definitions of free and bound occurrences of variables are also standard, with $\ada,\ade$ acting as quantifiers along with $\cla,\cle$.

{\bf Trees of formulas} are the elements of the smallest set of expressions such that:\vspace{-3pt}

\begin{itemize}
\item The empty expression is a tree of formulas, said to be {\bf empty}.\vspace{-3pt}
\item Every formula is a ({\bf nonempty}) tree of formulas.\vspace{-3pt}
\item If ${\cal T}$ and ${\cal S}$ are nonempty trees of formulas, then so is $({\cal T})\circ ({\cal S})$. Here parentheses can be omitted around $\cal T$ and $\cal S$ if they are formulas.\vspace{-3pt}
\end{itemize}

As in the case of trees of games, the {\bf yield} of a tree of formulas is the sequence of formulas appearing in it, in the left to right order. The yield of the empty tree is empty. We will often write a tree $\cal T$ of formulas as ${\cal T}(E_1,\ldots,E_n)$ to indicate that $E_1,\ldots,E_n$ is its yield. When $n=0$, ${\cal T}(E_1,\ldots,E_n)$ is the empty tree.

A {\bf sequent} is ${\cal T}(E_1,\ldots,E_n)\fintimpl F$ ($n\geq 0$), where ${\cal T}(E_1,\ldots,E_n)$, called the {\bf antecedent},  is a tree of formulas, and $F$, called the {\bf succedent}, is a  formula. When $n=0$, such a sequent simply looks like $\fintimpl F$, and is said to be an {\bf empty-antecedent sequent}.

Sometimes a formula $F$ will be represented as $F(x_1,\ldots,x_n)$, where the $x_i$ are pairwise distinct variables. 
When doing so, we do not necessarily mean that each such $x_i$ has a free occurrence in $F$, or that every variable occurring free in $F$ is among $x_1,\ldots,x_n$. In the context set by the above representation, $F(t_1,\ldots,t_n)$ will mean the result of replacing, in $F$, each free occurrence of each $x_i$ ($1\leq i\leq n$) by term $t_i$. 
The same notational conventions also apply to sequents and trees of formulas.

An {\bf interpretation}\footnote{The concept of an interpretation in CL is usually more general than the present one, where interpretations in our present sense are called  {\bf perfect}. But here we omit the word ``perfect'' as we do not consider any nonperfect interpretations, anyway.} is a function $^*$ that sends each
 $n$-ary function symbol $f$ to a function \(f^*:\ \{0,1,2,\ldots\}^n$ $\rightarrow \{0,1,2,\ldots\}\); it also sends each 
 $n$-ary  predicate symbol $p$ to an $n$-ary elementary game $p^*(x_1,\ldots,x_n)$ which does not depend on any variables other than $x_1,\ldots,x_n$; the additional condition required to be satisfied by $^*$ is that $=^*$ is an equivalence relation on $\{0,1,2,\ldots\}$ preserved by $f^*$ for each function symbol $f$, and respected by $p^*$ for each nonlogical predicate symbol $p$.\footnote{More commonly classical logic simply treats  $=$ as the identity predicate. That treatment of $=$, however, is known to be equivalent --- in every respect relevant for us --- to our present one. Namely, the latter turns into the former by seeing 
any two $=^*$-equivalent constants  as two different {\em names} of the same {\em object} of the universe, as ``Evening Star'' and ``Morning Star'' are.}    

The above uniquely extends to  a  mapping that sends each term $t$ to a function $t^*$, each formula $F$ to a game $F^*$, each nonempty tree $\cal T$ of formulas to a tree ${\cal T}^*$ of games, and each sequent $X$ to a game $X^*$ by stipulating that:\vspace{-5pt} 
\begin{enumerate}
\item $x^*=x$ (any variable $x$).\vspace{-5pt}
\item $c^*=c$ (any constant $c$).\vspace{-5pt}
\item Where $f$ is an $n$-ary function symbol and $t_1,\ldots,t_n$ are terms, $\bigl(f(t_1,\ldots,t_n)\bigr)^*=f^*(t_{1}^{*},\ldots,t_{n}^{*})$.\vspace{-6pt}
\item Where $p$ is an $n$-ary  predicate symbol  and $t_1,\ldots,t_n$ are terms, $\bigl(p(t_1,\ldots,t_n)\bigr)^*=p^*(t_{1}^{*},\ldots,t_{n}^{*})$.\vspace{-6pt} 
\item $^*$ commutes with $\circ$, in the sense that $(E\circ F)^*=E^*\circ F^*$.\vspace{-5pt}
\item $^*$ commutes with $\fintimpl$ when the antecedent is not empty: $({\cal T}\fintimpl F)^*={\cal T}^*\fintimpl F^*$.\vspace{-5pt}
\item An empty-antecedent sequent is simply understood as its succedent: $(\fintimpl F)^*=F^*$.\vspace{-5pt}
\item $^*$ commutes with all remaining operators:  $\twg^*=\twg$, $(\gneg F)^*=\gneg F^*$, $(E\mlc F)^*=E^{*}\mlc F^{*}$, $(\ada x F)^*=\ada x(F^*)$, etc.\vspace{-5pt} 
\end{enumerate}

When $O^*=P$ (whether $O$ be a predicate symbol, function  symbol, formula or sequent), we say that $^*$ {\bf interprets} $O$ as $P$. If we say ``{\bf $O$ under interpretation $^*$}'', what we mean is $O^*$.

When a given formula is represented as $F(x_1,\ldots,x_n)$, we will typically write $F^*(x_1,\ldots,x_n)$ instead of 
$\bigl(F(x_1,\ldots,x_n)\bigr)^*$.

Let $X$ be a sequent. For an   HPM, EPM or BMEPM $\cal M$, we say that $\cal M$ is a {\bf uniform solution}
for $X$ iff ${\cal M}\models {X}^*$ for every interpretation $^*$. 
And we say that $X$ is {\bf uniformly valid} iff there is a uniform solution for $X$.

\section{Some closure properties of computability}\label{ss7}

In this section we establish certain, mostly rather obvious yet important, closure properties for the computability of static games. While these results can be stated and proven in  more general forms than presented here, for simplicity we restrict them to games expressible in the formal language of Section \ref{ss6}.

By a ``{\bf rule}'' in this section we mean an $(n+1)$-ary relation $\cal R$ on sequents, the instances of which are schematically written as  
\[\frac{X_1\hspace{20pt}\ldots\hspace{20pt}X_n}{X_0},\]
where $X_1,\ldots,X_n$ are (metavariables for) sequents called the {\bf premises}, and $X_0$ is a sequent called the {\bf conclusion}. 
Whenever ${\cal R}(X_1,\ldots,X_n,X_0)$ holds, we say that $X_0$ {\bf follows from} $X_1,\ldots,X_n$ by $\cal R$.

We say that such a rule $\cal R$ is {\bf sound} iff it preserves computability, in the sense that, whenever each premise of a given instance of the rule is computable under a given interpretation, so is the conclusion (under the same interpretation). And we say that ${\cal R}$ is {\bf uniform-constructively sound}  iff there is an effective procedure that takes any instance $(X_1,\ldots,X_n,X_0)$ of the rule, any BMEPMs  ${\cal M}_1,\ldots,{\cal M}_n$ and returns a BMEPM ${\cal M}_0$ such that, for any interpretation $^*$ and valuation $e$,  whenever  ${\cal M}_1\models_e X_{1}^{*},\ldots,{\cal M}_n\models_e X_{n}^{*} $, we have ${\cal M}_0\models_e X_{0}^{*}$.

Below we prove several uniform-constructive soundness results. The proofs will be limited to showing how to construct ${\cal M}_0$ from $(X_1,\ldots,X_n,X_0)$ and ${\cal M}_1,\ldots,{\cal M}_n$. It will be immediately clear from our descriptions of the ${\cal M}_0$s that they can be constructed effectively (so that the closure is ``constructive''), and that their work in no way depends on an interpretation $^*$ applied to the sequents involved (so that the closure is ``uniform''). Since an interpretation $^*$ is typically irrelevant in such proofs, we will often omit it and write simply $F$ or $X$ where, strictly speaking, $F^*$ or $X^*$ is meant. That is, we identify formulas or sequents with the games into which they turn once an interpretation is applied to them. Similarly, a valuation $e$ is often irrelevant, and we write $F$ or $X$ where, strictly speaking, $e[E^*]$ or $e[X^*]$ is meant.  

We agree that in this and subsequent sections, the expression \[E_1,\ldots, E_n\fintimpl F\] is an abbreviation of 
\[E_1\circ E_2\circ E_3\circ\ldots E_{n-1}\circ E_n\fintimpl F\] which, in turn, is an abbreviation of  \[E_1\circ(E_2\circ (E_3\circ\ldots (E_{n-1}\circ E_n)\ldots))\fintimpl F.\]
When $n=0$,  the sequent $E_1,\ldots, E_n\fintimpl F$ is just $\fintimpl F$. The expression ``$E_1,\ldots,E_n$'' will often be abbreviated as $\vec{E}$. 

Throughout this section, $n$ is a natural number. Remember that we  write ${\cal T}(E_1,\ldots,E_n)$ for an arbitrary tree $\cal T$ of formulas with yield $E_1,\ldots,E_n$. Again, when $n=0$, ${\cal T}(E_1,\ldots,E_n)$ is the empty tree. When ${\cal T}(E_1,\ldots,E_n)=E_1\circ\ldots\circ E_n$ or $\cal T$ is empty, we say that $\cal T$ is {\bf standard}. 

 \subsection{Standardization and Destandardization} Standardization and Destandardization are the following two rules, respectively:  
\[\frac{{\cal T}(E_1,\ldots,E_n)\fintimpl F}{E_1,\ldots,E_n\fintimpl F}\hspace{100pt}\frac{E_1,\ldots,E_n\fintimpl F}{{\cal T}(E_1,\ldots,E_n)\fintimpl F}\]

Standardization thus allows us to replace any tree of formulas in the antecedent of the premise by a standard tree with the same yield, and Destandardization does the opposite.  

\begin{proposition}\label{may4a}
Both Standardization and Destandardization are  uniform-constructively sound.
\end{proposition}
\begin{proof} Here we only consider Standardization, with the case of Destandardization being fully symmetric. 

Assume ${\cal M}_1$ is a BMEPM that wins the premise under a given interpretation on a given valuation.  Note that the premise and the conclusion of the rule are ``essentially the same'', with the minor difference that the address of each leaf $E_i$ in the premise may be different from the address of the corresponding leaf $E_i$ in the conclusion. Of course, one can easily figure out a bijection $f$ that translates the addresses. We construct ${\cal M}_0$ by letting it be a machine that simulates ${\cal M}_1$, imagining that the valuation spelled on the valuation tape of the latter is the same as the valuation spelled on its own valuation tape. In this simulation, ${\cal M}_0$ acts (grants permissions, makes moves) in the same way as ${\cal M}_1$, only translates any move $w.\alpha$ or $\col{w}$ by ${\cal M}_1 $ into $f(w).\alpha$ or $\col{f(w)}$ before mimicking it in the real play. Similarly, an environment's move $w.\alpha$ in the real play it translates into $f^{-1}(w).\alpha$ before feeding it into the simulated play. Replicative moves, of course, change $f$ (including its domain), but correspondingly updating 
$f$ each time is no problem. The only minor complication is related to the fact that, when the environment makes a move $u.\alpha$ in the real play, or ${\cal M}_1$ makes move $u.\alpha$ in the simulated play, $u$ may not necessarily be the address of a leaf of the current antecedental tree, but rather a prefix of some of such addresses. For instance, this is the case with the 7th move of Example \ref{ex37}. Let, in this case, $w_1,\ldots,w_k$ be all the addresses of the leaves of the current antecedental tree such that $u$ is a prefix of them. Note that then, the effect of the single move $u.\alpha$ is the same as the effect of $k$ consecutive moves $w_1.\alpha,\ldots,w_k\alpha$. So, ${\cal M}_0$ interprets the single move $u.\alpha$ as the block $w_1.\alpha,\ldots,w_k.\alpha$ before feeding it into the simulated play or mimicking it in the real play.  Remembering our assumption that ${\cal M}_1$ wins the premise, it is not hard to see that ${\cal M}_0$ wins  the conclusion.
\end{proof}

\subsection{Exchange} Exchange is the following rule:
\[\frac{\vec{E},H,G,\vec{K}\fintimpl F}{\vec{E},G,H,\vec{K}\fintimpl F}\]

\begin{proposition}\label{may4b}
Exchange is  uniform-constructively sound.
\end{proposition}
\begin{proof} The idea here is the same as in the proof of the previous proposition. ${\cal M}_0$ plays exactly as ${\cal M}_1$, only correspondingly adjusting the addresses, and occasionally interpreting a single move as a block of moves. 
\end{proof}

\subsection{Weakening}
Weakening is the following rule:  
\[\frac{\vec{E}\fintimpl F}{\vec{E},\vec{K}\fintimpl F}\]

\begin{proposition}\label{may4c}
Weakening is  uniform-constructively sound.
\end{proposition}
\begin{proof} Assume ${\cal M}_1$ is a BMEPM that wins the premise. We construct ${\cal M}_0$ by letting it be a machine that
plays exactly as ${\cal M}_1$ does, by simulating the latter. Only, ${\cal M}_0$ ignores any moves that the environment  makes within the $\vec{K}$ part of the antecedent, as if that part simply did not exist. In addition, if $\vec{E}$ is empty (while $\vec{K}$ is not), ${\cal M}_0$ adds the prefix ``$S.$'' when mimicking moves made by ${\cal M}_1$ in the succedent of the premise, and deletes the same prefix when feeding back into the simulated play the environment's moves in the succedent of the conclusion. Obviously, ${\cal M}_0$ wins the game. \end{proof}

\subsection{Contraction}

Contraction is the following rule:  
\[\frac{\vec{E},G,G\fintimpl F}{\vec{E},G\fintimpl F}\]

\begin{proposition}\label{may4d}
Contraction is  uniform-constructively sound.
\end{proposition}
\begin{proof} Again, assume ${\cal M}_1$ is a BMEPM that wins the premise. We construct ${\cal M}_0$ by letting it be a machine that makes a replicative move which brings  the conclusion down to the premise, and then continues as ${\cal M}_1$.   \end{proof}

\subsection{Instantiation}

Instantiation is the following rule:  
\[\frac{\vec{E}(x)\fintimpl F(x)}{\vec{E}(c)\fintimpl F(c)}\]
Here $x$ is any variable, $c$ is any constant, and the conclusion is the result of replacing, in the premise, all free occurrences of $x$ by $c$.

\begin{proposition}\label{may4e}
Instantiation is  uniform-constructively sound.
\end{proposition}
\begin{proof} Assume ${\cal M}_1$ is a BMEPM that wins the premise. We let ${\cal M}_0$ be the machine that, with a valuation $e$ spelled on its valuation tape, plays (by simulating ${\cal M}_1$) as ${\cal M}_1$ would play with valuation $e'$ on its valuation tape, where $e'$ is the valuation that sends $x$ to $c$ and agrees with $e$ on all other variables. \end{proof}

\subsection{Cut} Unlike all previous rules, Cut is a rule with two premises:
\[\frac{\vec{E}\fintimpl F\hspace{30pt}\vec{K},F\fintimpl G}{\vec{E},\vec{K}\fintimpl G}\]

\begin{theorem}\label{april22}
Cut is uniform-constructively sound.  
\end{theorem}

\begin{proof} A detailed technical proof of this theorem could be lengthy, and here we essentially only give a proof idea (as we, in fact, did for all the previous propositions of this section as well). The terminology we will be using below is even more relaxed than earlier. Specifically, it should be pointed out that to what we may refer as ``$F$'' (or ``$\vec{E}$'', etc.), strictly speaking, typically will not be the original $F$ (or $\vec{E}$, etc.), but rather the game to which this component has evolved by the time of the context of such a reference. 

Assume ${\cal M}_1$ and ${\cal M}_2$ are two BMEPMs that win $\vec{E}\fintimpl F$ and  $\vec{K},F\fintimpl G$, respectively. 

Let us first consider the simple case when the two machines never make any replicative moves in the antecedents of the corresponding games. We let ${\cal M}_0$ be the following BMEPM. Its work on a valuation $e$ consists of infinitely many cycles. Each cycle consists in simulating, in parallel, the two machines ${\cal M}_1$ and ${\cal M}_2$ with  the same $e$ on their valuation tapes. The simulation of either machine continues until it grants permission. ${\cal M}_0$ records all moves made by the simulated machines. Once both machines have granted permission, ${\cal M}_0$ halts simulation, and all moves made by the two machines  in the $\vec{E}$ , $\vec{K}$ and $G$ components of the premises it mimics in the same components of the conclusion which is played in the real play (after correspondingly readjusting the addresses of the leaves if necessary, and also perhaps interpreting certain moves as whole blocks of moves, as we have already seen in the previous proofs). Then ${\cal M}_0$ grants permission in the real play, and records the environment's response, which should be some (maybe none) moves within the above components. Now ${\cal M}_0$ resumes simulation, feeding these moves back to the corresponding machines as imaginary responses to their granting permission. In addition, it also feeds back to ${\cal M}_1$ all moves made by ${\cal M}_2$ within the $F$ component of the second premise, and vice versa: feeds back to ${\cal M}_2$ all moves made by ${\cal M}_1$ within the $F$ component of the first premise. Here the cycle ends, and the next cycle does the same, starting (continuing) simulation of the two machines until they both grant permission, and so on. With some thought and keeping in mind the crucial fact that the games we are talking about are static, one can see that ${\cal M}_0$ wins $\vec{E},\vec{K}\fintimpl G$. 

But, again, in the above simplified scenario we did not account for the possibility of replicative moves that ${\cal M}_1$ and/or ${\cal M}_2$ may make. If a subcomponent of $\vec{E}$ is replicated (by ${\cal M}_1$) or a subcomponent of $\vec{K}$ is replicated (by ${\cal M}_2$), ${\cal M}_0$ can replicate the same subcomponent in the conclusion, and otherwise continue playing according to the earlier scenario. This case thus does not really create any complications.  The case of ${\cal M}_2$ replicating the $F$ component, thus turning the game that it plays into $\vec{K},F,F\fintimpl G$,  is more serious though. In this case, ${\cal M}_0$ needs to split its simulation of ${\cal M}_1$ into two simulations (continuations), let us call them $\#1$ and $\#2$. It also correspondingly replicates the whole $\vec{E}$ part of the conclusion in the real play, which --- modulo Exchange and Standardization --- will now look like $\vec{E},\vec{E},\vec{K}\fintimpl G$.\footnote{Remember what was said at the beginning of the proof about our relaxed terminology: of course, neither $F$ nor $\vec{E}$ nor $\vec{K}$ nor $G$ will necessarily be exactly the same as the corresponding original subgames, but rather be whatever games they have been brought down to by now.} From now on, the moves made in the $F$ component by ${\cal M}_1$ in simulation $\#1$ (resp. $\#2$) will be fed back into the first (resp. second) copy of $F$ in the game $\vec{K},F,F\fintimpl G$ played by the simulated ${\cal M}_2$, and vice versa: 
the moves made by ${\cal M}_2$ within the first (resp. second) $F$ component of that game will be fed back into simulation $\#1$ (resp. $\#2$) of ${\cal M}_1$.  Similarly, ${\cal M}_0$ will be mimicking the moves made by ${\cal M}_1$ in simulation $\#1$ (resp. $\#2$) within the $\vec{E}$ component as moves in the first (resp. second) copy of $\vec{E}$ in $\vec{E},\vec{E},\vec{K}\fintimpl G$, and vice versa: feeding the environment's moves made in the first (resp. second) copy of $\vec{E}$ in $\vec{E},\vec{E},\vec{K}\fintimpl G$ into simulation $\#1$ (resp. $\#2$) of ${\cal M}_1$. 

Subsequent replications by ${\cal M}_2$ of any given copy of $F$ will be handled in a similar way, correspondingly further increasing the number of various parallel simulations of ${\cal M}_1$ and the number of copies of $\vec{E}$ in the antecedent of the conclusion.    
\end{proof}

\section{Logic $\cltw$}\label{ss8}

The purpose of the deductive system $\cltw$ that we construct in this section is to axiomatize the set of uniformly valid sequents. In view of the closure under Standardization and Destandardization proven in the previous section, limiting sequents to those with standard antecedents does not yield any real loss of expressive power,\footnote{Some readers may be wondering why we did not then try to define dfb-reduction as an operation applied to sequences rather than trees of games (in the antecedent) in the first place. There are reasons, discernable only to an expert on computability logic. Among such reasons is that any attempt to replace trees with sequences would encounter the problem of violating the (crucially important) static property of games. This is so because the effect of a move by $\oo$ could depend on whether it was made before or after a replicative move by $\pp$,  and it could thus change when one passes from a run to a $\oo$-delay of it.}  and this is what we do from now on: the language of $\cltw$ exclusively deals with such sequents. Correspondingly, from now on, when we say ``{\bf sequent}'', we exclusively mean one of the form 
\[E_1,\ldots, E_n\fintimpl F\]
($n\geq 0$). 
For safety and also without loss of expressive power, we further agree that, from now on, the formulas or sequents that we consider may not contain both bound and free occurrences of the same variable. Other than these minor revisions, the language of $\cltw$ is the same as the one introduced in Section \ref{ss6}.  
 
Our formulation of $\cltw$ relies on some terminology and notation explained below.

\begin{enumerate}
\item A \gj{surface occurrence} of a subformula is an occurrence that is 
not in the scope of any choice operators. 
\item A formula not containing choice operators --- i.e., a formula of the classical language --- is said to be \gj{elementary}. 
\item A sequent is \gj{elementary} iff all of its formulas are so. 
\item The \gj{elementarization} \[\elz{F}\] of a formula $F$ is the result of replacing
in $F$ all $\add$- and $\ade$-subformulas by $\tlg$, and all $\adc$- and $\ada$-subformulas by $\twg$. Note that $\elz{F}$ is (indeed) an elementary formula.
\item The {\bf elementarization} of a sequent 
$G_1,\ldots,G_n\fintimpl F$ is the elementary formula \[\elz{G_1}\mlc\ldots\mlc \elz{G_n}\mli \elz{F}.\] 
\item A sequent  is said to be \gj{stable} iff its elementarization is classically valid.  By ``classical validity'', in view of G\"{o}del's completeness theorem,  we mean provability in classical first-order calculus with constants, function symbols and $=$, where $=$ is treated as the logical {\em identity} predicate (so that, say, $x=x$, $x=y\mli (E(x)\mli E(y))$, etc. are provable).
\item We will be using the notation \[F[E]\] to mean a formula $F$ together with some (single) fixed  surface occurrence of a subformula $E$. Using this notation sets a context, in which $F[H]$ will mean the result of replacing in $F[E]$ the (fixed) occurrence of $E$ by $H$.  Note that here we are talking about some {\em occurrence} of $E$. Only that occurrence gets replaced when moving from $F[E]$ to $F[H]$, even if the formula also had some other occurrences of $E$.
\end{enumerate}

\begin{center}
\begin{picture}(100,30)

\put(0,10){\bf THE RULES OF $\cltw$}

\end{picture}
\end{center}

$\cltw$ has the six rules listed below, with the following additional conditions/explanations: 
\begin{enumerate}
\item In $\adc$-Choose and $\add$-Choose, $i\in\{0,1\}$.
\item In $\ada$-Choose and $\ade$-Choose,  $t$ is either a constant or a variable with no bound occurrences in the premise, and $H(t)$ is the result of replacing by $t$ all free occurrences of $x$ in $H(x)$ (rather than vice versa).
\end{enumerate}
\begin{center}
\begin{picture}(287,70)

\put(214,50){\bf $\add$-Choose}
\put(212,30){$\vec{G}\ \fintimpl\  F[H_i]$}
\put(200,22){\line(1,0){78}}
\put(200,8){$\vec{G}\ \fintimpl \ F[H_0\add H_1]$}

\put(32,50){\bf $\adc$-Choose}
\put(12,30){$\vec{G},\ E[H_i],\ \vec{K}\  \fintimpl \ F$}
\put(0,22){\line(1,0){113}}
\put(0,8){$\vec{G},\ E[H_0\adc H_1], \ \vec{K}\ \fintimpl\ F$}

\end{picture}
\end{center}

\begin{center}
\begin{picture}(287,70)

\put(31,50){\bf $\ada$-Choose}
\put(10,30){$\vec{G},\ E[H(t)],\ \vec{K}\ \fintimpl\ F$}
\put(0,22){\line(1,0){113}}
\put(0,8){$\vec{G},\ E[\ada xH(x)],\ \vec{K}\ \fintimpl\ F$}

\put(216,50){\bf $\ade$-Choose}
\put(208,30){$\vec{G}\ \fintimpl\  F[H(t)]$}
\put(200,22){\line(1,0){78}}
\put(200,8){$\vec{G}\ \fintimpl \ F[\ade x H(x)]$}

\end{picture}
\end{center}

\begin{center}
\begin{picture}(74,70)

\put(12,50){\bf Replicate}
\put(8,8){$\vec{G},E,\vec{K}\fintimpl F$}
\put(0,22){\line(1,0){69}}
\put(0,30){$\vec{G},E,\vec{K},E\fintimpl F$}
\end{picture}
\end{center}

\begin{center}
\begin{picture}(300,70)
\put(140,50){\bf Wait}
\put(0,30){$X_1,\ldots,X_n$}
\put(0,22){\line(1,0){45}}
\put(55,20){($n\geq 0$), where all of the following five conditions are satisfied:}
\put(20,8){$Y$}
\end{picture}
\end{center}

\begin{enumerate}
\item {\bf $\adc$-Condition:}  Whenever $Y$ has the form $\vec{G}\fintimpl F[H_0\adc H_1]$, both of the sequents $\vec{G}\fintimpl F[H_0]$ and 
$\vec{G}\fintimpl F[H_1]$ are among $X_1,\ldots,X_n$.
\item {\bf $\add$-Condition:} Whenever $Y$ has the form $\vec{G},E[H_0\add H_1],\vec{K}\fintimpl F$, both of the sequents $\vec{G},E[H_0],\vec{K}\fintimpl F$ and 
$\vec{G},E[H_1],\vec{K}\fintimpl F$ are among $X_1,\ldots,X_n$.
\item {\bf $\ada$-Condition:} Whenever $Y$ has the form $\vec{G}\fintimpl F[\ada xH(x)]$, for some variable $y$ not occurring in $Y$, the sequent  $\vec{G}\fintimpl F[H(y)]$ is among  $X_1,\ldots,X_n$. Here and below, $H(y)$ is the result of replacing by $y$ all free occurrences of $x$ in $H(x)$ (rather than vice versa).
\item {\bf $\ade$-Condition:} Whenever $Y$ has the form $\vec{G},E[\ade xH(x)],\vec{K}\fintimpl F$, for some variable $y$ not occurring in $Y$, the sequent  $\vec{G},E[H(y)],\vec{K}\fintimpl F$ is among  $X_1,\ldots,X_n$.
\item {\bf Stability Condition:} $Y$ is stable.
\end{enumerate}

A {\bf $\cltw$-proof} of a sequent $X$ is a sequence $X_1,\ldots,X_n$ of sequents, with $X_n=X$, such that, each $X_i$ follows  by one of the rules of $\cltw$ from some (possibly empty in the case of Wait, and certainly empty in the case of $i=1$) set $\cal P$ of premises such that ${\cal P}\subseteq \{X_1,\ldots, X_{i-1}\}$.
When a $\cltw$-proof of $X$ exists, we say that $X$ is {\bf provable} in $\cltw$, and write $\cltw\vdash X$.   

\begin{example}\label{apr5}
 In this example, $+$ is a binary function symbol. We write $t_1+t_2$ instead of $+(t_1,t_2)$. The following sequence of sequents is a $\cltw$-proof (of its last sequent):\vspace{7pt}

\noindent 1.  $\cla x(x+0=x)\fintimpl w+0=w$ \ \ { Wait:} (no premises) \vspace{3pt}

\noindent 2.  $\cla x(x+0=x)\fintimpl \ade z(w+0=z)$ \ \ { $\ade$-Choose: 1}\vspace{3pt} 

\noindent 3.  $\cla x(x+0=x)\fintimpl \ada y\ade z(y+0=z)$ \ \ { Wait: 2} 
\end{example}

\begin{example} The following is a $\cltw$-proof:\vspace{7pt}

\noindent 1. $\cla x \cla y\cla z\bigl(p(x,y)\mlc p(y,z)\mli q(x,z)\bigr),\   p(w,v), \ p(u,w)\ \fintimpl \ q(u,v)$ \ \ {Wait:} \vspace{3pt}

\noindent 2. $\cla x \cla y\cla z\bigl(p(x,y)\mlc p(y,z)\mli q(x,z)\bigr),\  p(w,v), \ p(u,w)\ \fintimpl \ \ade yq(u,y)$ \ \ {$\ade$-Choose: 1}\vspace{3pt}

\noindent 3. $\cla x \cla y\cla z\bigl(p(x,y)\mlc p(y,z)\mli q(x,z)\bigr),\  \ade y p(w,y), \ p(u,w)\ \fintimpl \ \ade yq(u,y)$ \ \ { Wait: 2}\vspace{3pt}

\noindent 4. $\cla x \cla y\cla z\bigl(p(x,y)\mlc p(y,z)\mli q(x,z)\bigr),\ \ada x\ade y p(x,y), \ p(u,w)\ \fintimpl \ \ade yq(u,y)$ \ \ { $\ada$-Choose: 3}\vspace{3pt}

\noindent 5. $\cla x \cla y\cla z\bigl(p(x,y)\mlc p(y,z)\mli q(x,z)\bigr),\ \ada x\ade y p(x,y), \ \ade y p(u,y)\ \fintimpl \ \ade yq(u,y)$ \ \ { Wait: 4}\vspace{3pt}

\noindent 6. $\cla x \cla y\cla z\bigl(p(x,y)\mlc p(y,z)\mli q(x,z)\bigr),\ \ada x\ade y p(x,y),\ \ada x\ade y p(x,y)\ \fintimpl \ \ade yq(u,y)$  \ \ { $\ada$-Choose: 5}\vspace{3pt}

\noindent 7. $\cla x \cla y\cla z\bigl(p(x,y)\mlc p(y,z)\mli q(x,z)\bigr),\ \ada x\ade y p(x,y)\ \fintimpl \ \ade yq(u,y)$ \ \ {Replicate: 6}\vspace{3pt}

\noindent 8. $\cla x \cla y\cla z\bigl(p(x,y)\mlc p(y,z)\mli q(x,z)\bigr),\ \ada x\ade y p(x,y)\ \fintimpl \ \ada x\ade yq(x,y)$ \ \ { Wait: 7}
\end{example}

\begin{example}\label{apr5a} In this $\cltw$-proof, $+$ is as in Example \ref{apr5}, and $'$ is a unary function symbol. We  write $t'$ instead of $'(t)$.\vspace{7pt}

\noindent 1. $\cla x\cla y\bigl(x+y'=(x+y)'\bigr), \    s=v',  \ u+w=v\fintimpl u+w'=s$ \ \ { Wait: }\vspace{3pt}

\noindent 2. $\cla x\cla y\bigl(x+y'=(x+y)'\bigr),  \  s=v', \   u+w=v\fintimpl \ade z (u+w'=z)$ \ \ {$\ade$-Choose: 1 }\vspace{3pt}

\noindent 3. $\cla x\cla y\bigl(x+y'=(x+y)'\bigr), \   \ade y(y=v'),  \ u+w=v\fintimpl \ade z (u+w'=z)$ \ \ { Wait: 2}\vspace{3pt}

\noindent 4. $\cla x\cla y\bigl(x+y'=(x+y)'\bigr), \ \ada x\ade y(y=x'),  \ u+w=v\fintimpl \ade z (u+w'=z)$ \ \ { $\ada$-Choose: 3}\vspace{3pt}

\noindent 5. $\cla x\cla y\bigl(x+y'=(x+y)'\bigr), \ \ada x\ade y(y=x'),  \ \ade z (u+w=z)\fintimpl \ade z (u+w'=z)$ \ \ { Wait: 4}\vspace{3pt}

\noindent 6. $\cla x\cla y\bigl(x+y'=(x+y)'\bigr), \ \ada x\ade y(y=x'), \ \ada y\ade z (y+w=z)\fintimpl \ade z (u+w'=z)$ \ \ {$\ada$-Choose: 5}\vspace{3pt}

\noindent 7. $\cla x\cla y\bigl(x+y'=(x+y)'\bigr), \ \ada x\ade y(y=x'), \ \ada y\ade z (y+w=z)\fintimpl \ada y\ade z (y+w'=z)$ \ \  {Wait: 6}
\end{example}

\section{The soundness and completeness of \cltw}\label{ssc}

Note that interpretations, as defined in this paper,  are nothing but first-order models (structures) in the classical sense with domain $\{0,1,2,\ldots\}$.  By G\"{o}del's completeness theorem, an (elementary) formula $F$ is valid (in the classical sense) iff it is true in every such model. This, in view of the classical behavior of $\gneg,\mlc,\mld,\cla,\cle$ when applied to elementary formulas, means that, for an elementary formula, validity in the classical sense means the same as uniform validity in our sense. We may implicitly rely on this fact in the sequel.

The following lemma can be verified by straightforward induction on the complexity of $F$:

\begin{lemma}\label{new1}
For any formula $F$, interpretation $^*$ and valuation $e$, $\win{F^*}{e}\emptyrun=\win{\elzi{F}^*}{e}\emptyrun$.
\end{lemma}

The above lemma easily extends from formulas to sequents:

\begin{lemma}\label{new1a}
For any sequent $X$, interpretation $^*$ and valuation $e$, $\win{X^*}{e}\emptyrun=\win{\elzi{X}^*}{e}\emptyrun$.
\end{lemma}

Below and elsewhere, by the {\bf length} of a $\cltw$-proof we mean the number of sequents appearing in the proof. 

\begin{lemma}\label{May2a} \ 
Let $X$ and $Y$ be arbitrary sequents such that one of the following four conditions is satisfied:\vspace{-3pt}
\begin{enumerate}
\item  $X$ has the form $\vec{E}\fintimpl F[H_0\adc H_1]$,  and $Y$ is 
$\vec{E}\fintimpl F[H_i]$ ($i=0$ or $i=1$).\vspace{-3pt}
\item  $X$ has the form $\vec{E}, G[H_0\add H_1], \vec{K}\fintimpl F$, and $Y$ is $\vec{E},G[H_i],\vec{K}\fintimpl F$ ($i=0$ or $i=1$).\vspace{-3pt}
\item  $X$ has the form $\vec{E}\fintimpl F[\ada xH(x)]$,  and $Y$ is $\vec{E}\fintimpl F[H(y)]$ for some variable $y$ not occurring in $X$.\vspace{-3pt}
\item  $X$ has the form $\vec{E},G[\ade xH(x)],\vec{K}\fintimpl F$,  and $Y$ is $\vec{E},G[H(y)],\vec{K}\fintimpl F$ for some variable $y$ not occurring in $X$.\vspace{-3pt}
\end{enumerate}

Then, any $\cltw$-proof $\cal P$ of $X$ of length $k$ can be effectively converted into a $\cltw$-proof $\cal Q$ of $Y$ of length $\leq k$. Furthermore, if $X$ is derived by Wait in $\cal P$, then the length of such a  $\cltw$-proof $\cal Q$ of $Y$ will be strictly less than $k$.
\end{lemma}

\begin{proof} Consider arbitrary sequents  $X$ and $Y$, with $X$ having a $\cltw$-proof $\cal P$ of length $k$. Using induction on $k$, the lemma can be proven by cases, depending on which of the four conditions of Lemma \ref{May2a} is satisfied by $X$ and $Y$.  

{\em Case 1:}  $X$ and $Y$ satisfy condition 1 of the lemma. Let us first consider the case when $X$ is derived in $\cal P$ by Wait. Note that then $Y$ is among its premises, and hence has a proof of length less than $k$. That very proof will be the sought proof $\cal Q$ of $Y$. Now suppose $X$ is derived by any of the other rules from a premise $U$. Note that $U$ has the same form $\vec{E}'\fintimpl F'[H_0\adc H_1]$ (some $\vec{E}',F'$) as $X$ has. Let $W$ be  $\vec{E}'\fintimpl F'[H_i]$. We can apply the induction hypothesis to $U$ and $W$ (in the role of $X$ and $Y$) and find that $W$ has a proof (which can be effectively constructed) of length less than $k$. But note that $Y$ follows from $W$ by the same rule as $X$ follows from $U$. So, $Y$ has a proof of length at most $k$.

{\em Case 2:}  $X$ and $Y$ satisfy condition 2 of the lemma. This is similar to the previous case. 

{\em Case 3:}  $X$ and $Y$ satisfy condition 3 of the lemma. First we consider the case when $X$ is derived in $\cal P$ by Wait. Note that then a certain sequent $Y'$, which is ``essentially the same'' as $Y$, is among the premises of $X$, and hence has a proof of length less than $k$. Here $Y'$ is ``essentially the same'' as $Y$ in the sense that it is either $Y$, or is different from $Y$ only in that it has a variable $z$ ($z\not=y$) where $Y$ has $y$. Obviously this is not a significant difference, and the proof of $Y'$ can be turned into a proof of $Y$ by just renaming variables. 
Such a proof of $Y$ will be the sought $\cal Q$. Now it remains to consider the case of  $X$ being derived by any of the other rules. Our reasoning in this case will be similar to the one given in Case 1.  

{\em Case 4:}  $X$ and $Y$ satisfy condition 4 of the lemma. Similar to the previous case. 
\end{proof}

Let us say that a sequent $Y$ is a {\bf $\oo$-development} of a sequent $X$ iff $X$ and $Y$ satisfy one of the four conditions of Lemma \ref{May2a}. And we say that $Y$ is a {\bf $\pp$-development} of  $X$ iff $X$ follows from $Y$ by one of the  rules  of $\cltw$ other than Wait. 

Now, we say that $Y$ is a {\bf transitive $\oo$-development} of $X$ iff there are sequents $Z_1,\ldots,Z_n$ ($n\geq 2$) where $Z_1=X$, $Z_n=Y$ and, for each $i$ with $1\leq i<n$, $Z_{i+1}$ is a $\oo$-development of $Z_i$.  
{\bf Transitive $\pp$-development} is defined similarly. 

\begin{lemma}\label{May2b} Let $X$ and $Y$ be any sequents.  

1. If $Y$ is a transitive $\oo$-development of $X$, \ $X$ has a $\cltw$-proof $\cal P$ of length $k$ and the last sequent  of the proof ($X$) is derived by Wait, then $Y$ has a $\cltw$-proof $\cal Q$ of length less than $k$. Such a $\cal Q$ can be effectively constructed from $\cal P$.

2. If $Y$ is a transitive $\pp$-development of $X$ and  $\cltw\not\vdash X$, then $\cltw\not\vdash Y$. 
\end{lemma}

\begin{proof} Clause 1 is an immediate corollary of Lemma \ref{May2a}. And clause 2 is straightforward. 
\end{proof}
 
\begin{theorem}\label{main}
$\cltw\vdash X$ iff $X$ is uniformly valid (any sequent $X$).
Furthermore:

{\bf Uniform-constructive soundness} There is an effective procedure that takes any $\cltw$-proof of any sequent $X$ and 
constructs a uniform solution for $X$.  
\end{theorem}

\begin{proof}  Consider an arbitrary sequent $X$. 
\vspace{5pt}

{\em Soundness:} Assume $X$ has a $\cltw$-proof ${\cal P}_X$. Pick an arbitrary interpretation $^*$. We want to (show how to) construct a $^*$-independent EPM ${\cal M}_X$ that wins 
$X^*$. As an induction hypothesis, we assume that, for any $\cltw$-proof ${\cal P}_Y$ of any sequent $Y$, as long as ${\cal P}_Y$ is shorter than ${\cal P}_X$, we know how to (effectively) construct a $^*$-independent EPM ${\cal M}_Y$ that wins $Y^*$.  In each case below, $e$ is the (arbitrary) valuation spelled on the valuation tape of the to-be-constructed machine ${\cal M}_X$. As was done before, we may omit either $e$ or $^*$ or both and, say, write simply $X$ instead of  $e[X^*]$.  The work of ${\cal M}_X$ depends on by what rule $X$ is derived in ${\cal P}_X$. 

{\em Case 1:} $X$ is derived from a premise $Y$ by $\add$-Choose. Then we let ${\cal M}_X$ be the machine that makes a move that brings $X$ down to $Y$. For instance, if $X$ is $\vec{K}\fintimpl E\mlc(F\mld (G\add H))$ and $Y$ is $\vec{K}\fintimpl E\mlc(F\mld G)$, then $S.1.1.0$ is such a move. After this, ${\cal M}_{X}$ continues as ${\cal M}_{Y}$ with the same valuation $e$ spelled on the valuation tape of the latter. By the induction hypothesis, ${\cal M}_{Y}$ wins $Y$, which obviously implies that ${\cal M}_X$ wins $X$.

{\em Case 2:} $X$ is derived from $Y$ by $\adc$-Choose. Then, as in the previous case,  ${\cal M}_{X}$ is the machine that makes a move that brings $X$ down to $Y$. For instance, if $X$ is $E, F\adc G, H\fintimpl K$ (i.e., $E\circ((F\adc G)\circ H)\fintimpl K$) and $Y$ is  $E, F, H \fintimpl K$ (i.e., $E\circ(F\circ H)\fintimpl K$), then $10.0$ is such a move. After this, ${\cal M}_X$ continues as ${\cal M}_Y$ with the same valuation $e$ spelled on the valuation tape of 
the latter. By the induction hypothesis, ${\cal M}_{Y}$ wins $Y$, which implies that ${\cal M}_X$ wins $X$.

{\em Case 3:} $X$ is derived from $Y$ by $\ade$-Choose. Then ${\cal M}_X$ is the machine that makes a move that brings $e[X]$ down to $e[Y]$. For instance, if $X$ is $\vec{K}\fintimpl E\mlc(F\mld \ade x H(x))$ and $Y$ is $\vec{K}\fintimpl E\mlc (F\mld H(t))$, then $S.1.1.c$ is such a move, where $c=t$ if $t$ is a constant, and $c=e(t)$ if $t$ is a variable (in the latter case the machine will have to read $c$ from its valuation tape).  After this move, ${\cal M}_X$ continues as ${\cal M}_Y$ with the same valuation $e$ spelled on the valuation tape of the latter. By the induction hypothesis, ${\cal M}_Y$ wins $Y$, which implies that ${\cal M}_X$ wins $X$.

{\em Case 4:} $X$ is derived from $Y$ by $\ada$-Choose. This case is similar to the previous one(s), and we can afford to omit details. 

{\em Case 5}:   $X$ is derived from $Y$ by Replicate. Note that Replicate is nothing but a simple combination of Contraction and Exchange. By the induction hypothesis, we have a machine ${\cal M}_Y$ winning the premise. Then, in view of Propositions \ref{may4b} and \ref{may4d}, we can construct the desired  machine ${\cal M}_X$ that wins the conclusion. 

{\em Case 6}:  $X$ is derived by Wait. Then ${\cal M}_{X}$  keeps granting permission until the environment makes a move. If such a move is never made, then the run that is generated is empty. As $X$ is derived by Wait, it is stable, so that $\elz{X}$ is classically valid. But then, in view of Lemma \ref{new1a}, the machine is the winner, no matter what valuation $e$ is spelled on its valuation tape. Suppose now the environment makes a move. We may assume that such a move is legal, or else the machine immediately wins. With a little thought, one can see that any legal move $\alpha$ by the environment brings the game down to $e'[Y]$ for a certain valuation $e'$ and a certain sequent $Y$ such that $Y$ is a transitive $\oo$-development of $X$. For example, if $X$ is $\vec{K}\fintimpl \ada x H(x)$ and $\alpha$ is $S.5$, then $Y$ can be chosen to be $\vec{K}\fintimpl H(y)$ for a variable $y$ not occurring in $X$, and then $e'$ would be the valuation that sends $y$ to $5$ and agrees with $e$ on all other variables, so that $e'[\vec{K}\fintimpl H(y)]$ is $e[\vec{K}\fintimpl H(5)]$, with the latter being the game to which $e[X]$ is brought down by the labmove $\oo S.5$. As another example, consider $X=\ade xE(x)\circ\ade xE(x)\fintimpl F$ and $\alpha = .5$ (so that the effect of $\alpha$ is making the move $5$ in both leaves of the antecedent). Then $Y$ can be chosen to be $E(y)\circ E(z)\fintimpl F$ for some fresh variables $y$ and $z$, in which case $e'$ would be the valuation that sends both $y$ and $z$ to $5$, and agrees with $e$ on all other variables.  Both $Y$ and $e'$ can be effectively found. Furthermore, by clause 1 of Lemma \ref{May2b}, $Y$ has a shorter proof $\cal Q$ than $X$ does, and $\cal Q$, too, can be effectively found. Then, by the induction hypothesis, an  EPM ${\cal M}_{Y}$ with ${\cal M}_{Y}\models Y$ can be further constructed effectively. So, our ${\cal M}_{X}$ now constructs ${\cal M}_{Y}$ and plays the rest of the game as ${\cal M}_{Y}$ would play with the valuation $e'$ spelled on its valuation tape. This obviously makes ${\cal M}_{X}$ successful.   
 \vspace{5pt}

{\em Completeness:}  Assume $\cltw\not\vdash X$. Here we describe a {\em counterstrategy} for $X$ in the form of an EPM $\cal C$ with an oracle for $\cltw$-provability. ``Counterstrategy'' means that $\cal C$ plays as EPMs generally do, with the difference that it plays in the role of $\oo$ rather than $\pp$. That is, the moves it makes get the label $\oo$, and the moves by its adversary get the label $\pp$. It will be seen that $\cal C$ is always the winner for an appropriately selected valuation $g$ and interpretation $^*$. In view of Lemma 20.8 of \cite{Jap03}, this implies that $X$ has no uniform solution. 

For a sequent $Y$ and valuation $e$, we say that $e$ is {\bf $Y$-distinctive} iff $e$ assigns different values to different free variables of $Y$, and all those values are also different from any constants occurring in $Y$. We fix $g$ as an arbitrary $X$-distinctive valuation, which is going to be the above-mentioned ``appropriately selected valuation''.
 
The work of $\cal C$, in a sense, is symmetric to the work of the machine ${\cal M}_X$ constructed in the proof of the soundness part. We define it recursively. At any time, it deals with a pair $(Y,e)$, where $Y$ is a $\cltw$-unprovable sequent and $e$ is a $Y$-distinctive valuation. The initial value of $Y$ is $X$, and the initial value of $e$ is the above $X$-distinctive valuation $g$, which we assume is the valuation spelled on the valuation tape of $\cal C$. How $\cal C$ acts on $(Y,e)$ depends on $Y$. 

If $Y$ is stable, then there should be a $\cltw$-unprovable sequent $Z$ satisfying one of the following conditions, for otherwise $Y$ would be derivable 
by Wait. Using its oracle for $\cltw$-provability, $\cal C$ finds and selects one such $Z$ (say, lexicographically the smallest one), and acts according to the corresponding prescription as given below. 

{\em Case 1:} $Y$ has the form $\vec{E}\fintimpl F[G_0\adc G_1]$, and $Z$ is $\vec{E}\fintimpl F[G_i]$ ($i=0$ or $i=1$). In this case, $\cal C$ makes a move that brings $Y$ down to $Z$, and calls itself on $(Z,e)$. 

{\em Case 2:} $Y$ has the form $\vec{E},F[G_0\adc G_1],\vec{K}\fintimpl H$, and $Z$ is $\vec{E},F[G_i],\vec{K}\fintimpl H$. This case is similar to the previous one. 
 
{\em Case 3:} $Y$ has the form $\vec{E}\fintimpl F[\ada xG(x)]$, and $Z$ is $\vec{E}\fintimpl F[G(y)]$, where $y$ is a variable not occurring in $Y$. In this case, $\cal C$ makes a move that brings $Y$ down to $\vec{E}\fintimpl F[G(c)]$ for some (say, the smallest) constant $c$ such that $c$ is different from any constant occurring in $Y$, as well as from any $e(z)$ where $z$ is a free variable of $Y$.  After this move, $\cal C$  calls itself on $(Z,e')$, where $e'$ is the valuation that sends $y$ to $c$ and agrees with $e$ on all other variables. 

{\em Case 4:} $Y$ has the form $\vec{E},F[\ada xG(x)],\vec{K}\fintimpl H$, and $Z$ is $\vec{E},F[G(y)],\vec{K}\fintimpl H$, where $y$ is a variable not occurring in $Y$. This case is similar to the previous one. 

Next, we consider the cases when $Y$ is not stable. Then ${\cal C}$  keeps granting permission until the environment makes a move. 

{\em Case 5}: If such a move is never made, then the run that is generated is empty. As $Y$ is not stable, $\elz{Y}$ is not classically valid. Also, as promised earlier and as it is easy to verify, $e$ is a $Y$-distinctive valuation and hence, of course, it is also $\elz{Y}$-distinctive.  It is a common knowledge from classical logic that, whenever a formula $F$ is invalid and $e$ is an $F$-distinctive valuation, $e[F]$ is false in some model. So, $e[\elz{Y}]$ is false in/under some model/interpretation $^*$. This, in view of Lemma \ref{new1a}, implies that $\win{Y^*}{e}\seq{}=\oo$ and hence $\cal C$ is the winner in the overall play of $X^*$ on $g$.

{\em Case 6}: Now assume the adversary makes a move. We may assume that such a move is legal, or else $\cal C$ immediately wins. With a little thought, one can see that any legal move $\alpha$ by the adversary will bring the game down to $e'[{Z}^*]$ for a certain $Z$-distinctive valuation $e'$ and sequent $Z$ such that $Z$ is a transitive $\pp$-development of $Y$. By clause 2 of Lemma \ref{May2b}, $\cltw\not\vdash Z$.  So, our machine just calls itself on 
$(Z,e')$.

It is clear that $\cal C$ wins as long as Case 6 occurs only finitely many times. And, if not, $\cal C$ again wins, as then the run is infinite. 
\end{proof}
   
\section{Natural deduction}\label{ss10}
We are going to construct a CL-based arithmetic in the form of a natural deduction system. This section explains the basic concepts and notational conventions for such systems in general, using the (slightly improved) Fitch style. 

Each natural deduction system will have a collection of {\bf rules} (of inference), and a {\bf deduction} in such a system will look something like the following:

\begin{center}
\begin{picture}(180,230)

\put(40,212){\bf Steps}

\put(110,212){\bf Justifications}

\put(32,8){\line(0,1){189}}
\put(58,22){\line(0,1){25}}
\put(58,69){\line(0,1){84}}
\put(83,83){\line(0,1){40}}
\put(32,171){\line(1,0){22}}
\put(83,111){\line(1,0){22}}
\put(58,141){\line(1,0){22}}
\put(58,36){\line(1,0){22}}
\put(40,190){$E_{1}$}
\put(40,175){$E_{2}$}
\put(40,160){$E_{3}$}
\put(65,145){$E_{4}$}
\put(65,130){$E_{5}$}
\put(90,115){$E_{6}$}
\put(90,100){$E_{7}$}
\put(90,85){$E_{8}$}
\put(65,70){$E_{9}$}
\put(40,55){$E_{10}$}
\put(65,40){$E_{11}$}
\put(65,25){$E_{12}$}
\put(40,10){$E_{13}$}

\put(0,190){$1.$}
\put(0,175){$2.$}
\put(0,160){$3.$}
\put(0,145){$4.1.$}
\put(0,130){$4.2.$}
\put(0,115){$4.3.1.$}
\put(0,100){$4.3.2.$}
\put(0,85){$4.3.3.$}
\put(0,70){$4.4.$}
\put(0,55){$5.$}
\put(0,40){$6.1.$}
\put(0,25){$6.2.$}
\put(0,10){$7.$}

\put(120,190){Hypothesis}
\put(120,175){Hypothesis}
\put(120,160){Rule$_3$: $\vec{n_3}$}
\put(120,145){Hypothesis}
\put(120,130){Rule$_5$: $\vec{n_5}$}
\put(120,115){Hypothesis}
\put(120,100){Rule$_7$: $\vec{n_7}$}
\put(120,85){Rule$_8$: $\vec{n_8}$}
\put(120,70){Rule$_9$: $\vec{n_9}$}
\put(120,55){Rule$_{10}$: $\vec{n_{10}}$}
\put(120,40){Hypothesis}
\put(120,25){Rule$_{12}$: $\vec{n_{12}}$}
\put(120,10){Rule$_{13}$: $\vec{n_{13}}$}

\end{picture}
\end{center}

Each $E_i$ here should be a formula (rather than sequent) of the language of $\cltw$, or whatever fragment of that language is considered in a given system. 
The above is a deduction of $E_{13}$ from the hypotheses $E_1$ and $E_2$. The steps $1$ through $7$ are those of the {\bf main (parent) deduction}. This deduction has two {\bf subdeductions} ({\bf children}). The first subdeduction comprises steps $4.1$ through $4.4$ (which can be compressed and, together, called step $4$ of the main deduction), and the second subdeduction comprises steps $6.1$ and $6.2$ (or, step $6$ of the main deduction). One of the steps of the first subdeduction, in turn, is itself a subdeduction, consisting of three steps. The vertical lines delimit (sub)deductions, and the horizontal line in each (sub)deduction separates the {\bf Hypotheses} from the rest of the steps. Each subdeduction thus may have its own hypotheses, in addition to those of its parents, grandparents, etc. On the other hand, the hypotheses of children do not count as those of parents, so that the deduction of $E_{13}$ that we see above is a deduction from (only) the hypotheses $E_1$ and $E_2$. 

Each formula in a deduction should come with a {\bf justification}. The justification for each hypothesis is the same: an indication that the step is a hypothesis (which, in fact, is redundant as the hypotheses can be recognized as the steps above the horizontal bar). And the justification for any other formula $E_i$ should be an indication of the rule Rule$_i$ by which the formula is obtained, and the step numbers $\vec{n_{i}}$ for the premises of the corresponding application of the rule.  
Here a rule is a relation of the type 
\[(\{\mbox{\em Formulas}\}\cup\{\mbox{\em Subdeductions}\})^n\ \rightarrow \{\mbox{\em Formulas}\}\]
($n$ may not necessarily be fixed, as in the LC rule of the following section). 
Such a rule has $n$ premises, and a justification should list each of those premises (identified by step numbers) together with the rule name. An important restriction is on what steps can serve as premises for a given application of a rule. First of all, each premise should be a step that occurs earlier than the given step. But not all earlier steps can be used. Specifically, the steps appearing inside earlier subdeductions that are no longer active (are ``closed'') cannot be referred to. Only the entire child subdeduction (rather than particular steps within it) can serve as a premise for a step in the parent deduction. For example, in the deduction of the above figure, the steps that $\vec{n_{12}}$ may include are limited to  $1$, $2$, $3$, $4$, $5$ and $6.1$. Note that here we have written $4$ as a reference to the entire  child subdeduction $4.1$-$4.4$. $\vec{n_{12}}$ cannot include any of the particular substeps $4.1$, $4.2$, $4.3$ or $4.4$ of $4$, or any of the substeps $4.3.1$, $4.3.2$ or $4.3.3$ of substep $4.3$. That is because all these steps are within already closed subdeductions. Similarly, the steps that   $\vec{n_{13}}$ can include are limited to $1$, $2$, $3$, $4$ (as a whole subdeduction), $5$ and $6$ (as a whole subdeduction). And the steps that $\vec{n_{9}}$ can include  are limited to $1$, $2$, $3$, $4.1$, $4.2$ and $4.3$; it cannot include $4.3.1$, $4.3.2$ or $4.3.3$. 

\section{Computability-logic-based arithmetic}\label{ss11}

There can be various reasonable systems of arithmetic based on computability logic, depending on what language we consider, what fragment of CL is taken as a logical basis, and what extra-logical rules and axioms are employed. The system $\ar$ offered in this section is one of them. Its language, whose formulas we call {\bf $\ar$-formulas} and whose sequents we call {\bf $\ar$-sequents}, is obtained from the language of $\cltw$ by removing all nonlogical predicate symbols (thus only leaving the logical predicate symbol $=$), all constants but $0$, and all but three function symbols, which are:\vspace{-3pt}
 
\begin{itemize}
\item $successor$, unary. We will write $t'$ for $successor(t)$.\vspace{-3pt}
\item $sum$, binary. We will write $t_1+t_2$ for $sum(t_1,t_2)$.\vspace{-3pt}
\item $product$, binary. We will write $t_1\times t_2$ for $product(t_1,t_2)$.\vspace{-3pt}
\end{itemize}

We shall also typically write $x\not= y$ for $\gneg x=y$.

The concept of an interpretation defined earlier can now be restricted to interpretations that are only defined on $=$, $'$, $+$ and $\times$, as the language has no other predicate or function  symbols. Of such interpretations, the {\bf standard interpretation} $^\dagger$ is the one that interprets $=$ as the identity relation, $'$ as the  successor ($x+1$) function, $+$ as the sum function,  and $\times$ as the product function. Where $F$ is a $\ar$-formula, the {\bf standard interpretation} 
of $F$ is the game $F^\dagger$, which we typically write simply as $F$. Similarly for $\ar$-sequents.

The axioms of $\ar$ are the $\cla$-closures of the following:\vspace{-3pt}

\begin{enumerate}
\item $0\not=x'$\vspace{-3pt}
\item $x'=y'\mli x=y$\vspace{-3pt}
\item $x+0=x$\vspace{-3pt}
\item $x+y'=(x+y)'$\vspace{-3pt}
\item $x\times 0=0$\vspace{-3pt}
\item $x\times y'=(x\times y)+x$\vspace{-3pt}
\item $\ada x\ade y(y=x')$\vspace{-3pt}
\item $F(0)\mlc \cla x(F(x)\mli F(x'))\mli \cla xF(x)$, where $F(x)$ is any elementary formula\vspace{-3pt}
\end{enumerate}

$\ar$ is a natural deduction system, and the above axioms are treated as rules of inference (in the sense of the previous section) that take no premises. In addition, $\ar$ has the following two rules:

\begin{center}
\begin{picture}(407,175)

\put(2,155){\bf Logical Consequence (LC)}
\put(68,134){$G_1$}
\put(71,116){$.$}
\put(71,121){$.$}
\put(71,126){$.$}
\put(68,101){$G_n$}
\put(71,83){$.$}
\put(71,88){$.$}
\put(71,93){$.$}
\put(65,66){\line(0,1){76}}
\put(68,68){$F$}
\put(58,70){\circle*{5}}
\put(0,48){where {\bf CL12} $\vdash G_1,\ldots,G_n\fintimpl F$}

\put(200,155){\bf Constructive Induction (CI)}
\put(268,134){$F(0)$}
\put(271,116){$.$}
\put(271,121){$.$}
\put(271,126){$.$}
\put(278,101){$F(x)$}
\put(275,96){\line(1,0){23}}
\put(281,80){$.$}
\put(281,85){$.$}
\put(281,90){$.$}
\put(278,65){$F(x')$}
\put(265,27){\line(0,1){113}}
\put(275,64){\line(0,1){45}}
\put(271,55){$.$}
\put(271,50){$.$}
\put(271,45){$.$}
\put(268,30){$\ada xF(x)$}
\put(258,32){\circle*{5}}
\put(220,10){where $x$ is a fresh variable}
\end{picture}
\end{center}

Here {\Large $\bullet$} indicates the conclusion of the rule, and whatever steps we see before the conclusion are the premises. Rule LC thus takes any $n$ ($n\geq 0$) premises $G_1,\ldots,G_n$ and, provided that  $\cltw\vdash G_1,\ldots,G_n\fintimpl F$, yields the conclusion $F$. The CI rule takes two premises: a formula of the form $F(0)$ and a subdeduction that has $F(x)$ as a single premise and $F(x')$ as the last step. The ``freshness'' requirement on $x$ means that $x$ should not occur anywhere earlier in the proof. This condition can be safely weakened by only requiring that $x$ does not occur in any earlier steps other than perhaps those within already closed subdeductions.  

Deductions in this system we call {\bf $\ar$-deductions}. 

A {\bf $\ar$-proof} of a formula $F$ is a $\ar$-deduction of $F$ from the empty set of hypotheses. 

$\ar$, however, can be used not only for proving formulas, but sequents as well. A {\bf $\ar$-proof} of a sequent $E_1,\ldots,E_n\fintimpl F$ is  a $\ar$-deduction of $F$ from hypotheses $E_1,\ldots,E_n$. 

An {\bf extended $\ar$-proof} of a sequent or formula $X$ is a $\ar$-proof of $X$ where to each application of the LC rule is attached, as an additional justification, a {\bf CL12}-proof of corresponding sequent, i.e. of the sequent whose antecedent consists of the  premises of the rule and whose succedent is the conclusion of the rule. 

When a $\ar$-proof of a sequent or formula $X$ exists, we say that $X$ is {\bf provable} in $\ar$ and write $\ar\vdash X$. 

Note that $\ar$-provability of a formula $F$ can be seen as a special case of the more general concept of provability of sequents. Specifically, it can be understood as $\ar$-provability of the empty-antecedent sequent $\fintimpl F$. 

$\ar$ is a computability-logic-based counterpart of the classical-logic-based {\bf Peano arithmetic} {\bf PA},  the best known and most studied applied formal theory of all. Formulas of {\bf PA} are nothing but elementary (choice-operator-free) formulas of the language of $\ar$. And the axioms of {\bf PA} are the same as those of $\ar$, only without the nonelementary Axiom 7. The latter says simply that the successor function is computable. And the CI rule, while formally being nothing but an ordinary induction principle with just $\cla$ replaced by $\ada$, in fact, provides for primitive recursion. A point worth emphasizing here is that, by just adding to {\bf PA} the very innocuous-looking Axiom 7 and rule CI, we ($\ar$, that is) get, via computability logic,  substantial information about computability. The information obtained in this way may not be new, but the fact that it is obtained by purely logical methods is a strong indication that computability logic can really do what it is intended to do. 

For typesetting considerations and without any danger of ambiguity, in the following examples we construct $\ar$-deductions without using vertical or horizontal bars. 

\begin{example}\label{apr5b}
 Proving $\ada x\ada y\ade z (y+x=z)$ in $\ar$:\vspace{7pt}

1. $\cla x(x+0=x)$ \ \ { Axiom 3}\vspace{3pt}

2. $\cla x\cla y\bigl(x+y'=(x+y)'\bigr)$ \ \ { Axiom 4}\vspace{3pt}

3. $\ada x\ade y(y=x')$ \ \ { Axiom 7}\vspace{3pt}

4. $\ada y\ade z(y+0=z)$ \ \ { LC: 1}\vspace{3pt} (Example \ref{apr5})

5.1. \ \ $\ada y\ade z (y+w=z)$ \ \ { Hypothesis}\vspace{3pt}

5.2. \ \ $\ada y\ade z (y+w'=z)$ \ \ { LC: 2, 3, 5.1}\vspace{3pt} (Example \ref{apr5a})

6. $\ada x\ada y\ade z (y+x=z)$ \ \ { CI: 4, 5}
\end{example}

Note that the classical counterpart $\cla x\cla y\cle z (y+x=z)$ of the above-proven formula is not very interesting ---  being logically valid, it carries no information whatsoever, and is true not only for $+$ but for any other function as well. As for  $\ada x\ada y\ade z (y+x=z)$, it is not trivial at all. In view of the forthcoming soundness theorem for $\ar$, its provability signifies that for any $x$ and $y$, a $z$ with $y+x=z$ not merely exists, but can also be algorithmically found; furthermore, such an algorithm, itself, can be constructed from a proof of the formula. So, Example \ref{apr5b} demonstrates the nontrivial (albeit not novel) fact that addition is a computable function; and the proof given there actually encodes an algorithm for computing this function. Similarly, the following example implies the computability of multiplication:

\begin{example}\label{j24b} Proving $\ada x\ada y\ade z (y\times x=z)$ in $\ar$:\vspace{7pt}

1. $\cla x(x\times 0=0)$ \ \ { Axiom 5}\vspace{3pt}

2. $\cla x\cla y\bigl(x\times y'=(x\times y)+x\bigr)$ \ \ { Axiom 6}\vspace{3pt}

3. $\ada x\ada y\ade z (y+x=z)$ \ \ (Example \ref{apr5b})\vspace{3pt}

4. $\ada y\ade z(y\times 0=z)$ \ \ { LC: 1}\vspace{3pt} (similar to Example \ref{apr5})

5.1. \ \ $\ada y\ade z (y\times w=z)$ \ \ {Hypothesis}\vspace{3pt}

5.2. \ \ $\ada y\ade z (y\times w'=z)$ \ \ { LC: 2, 3, 5.1}\vspace{3pt} (similar to Example \ref{apr5a})

6. $\ada x\ada y\ade z (y\times x=z)$ \ \ { CI: 4, 5}
\end{example}

In view of the well known deduction theorem, one of the many equivalent ways to define the provability of a formula $F$ in {\bf PA} is to say that $F$ is provable in {\bf PA} iff there are axioms $E_1,\ldots,E_n$ such that the formula $E_1\mlc\ldots\mlc E_n\mli F$ is provable in classical predicate calculus. Suppose the latter is indeed the case. Note that then the sequent $E_1,\ldots, E_n\fintimpl F$ is provable in $\cltw$ --- namely, derivable by Wait from the empty set of premises. Each $E_i$, being an axiom of ({\bf PA} and hence also of) $\ar$, is also provable in $\ar$. Then, by LC, $F$ is also provable. To summarize:

\begin{fact}\label{extention}
Every formula provable in {\bf PA} is also provable in $\ar$. 
\end{fact}

Thus, $\ar$ is an extension of {\bf PA}. A natural question to which at present we have no answer is whether $\ar$ is a conservative extension of {\bf PA}, that is, whether every {\em elementary} formula $F$ provable in $\ar$ is also provable in {\bf PA}. 

While, by G\"{o}del's incompleteness theorem, {\bf PA} fails to prove all true arithmetic formulas, {\bf PA} is still ``practically complete'', in the sense that it proves all reasonably simple (usually called ``elementary'', but not in our present sense of this word) true facts about natural numbers. Most readers are well familiar with {\bf PA} and have a good feel of what it can prove. This allows us to relax our proofs and, in presenting them, justify some ``steps'' (which, in reality, are sequences of many steps rather than single steps) by saying that the formula in the step is provable in {\bf PA}. 
 
\begin{example} The following sequence is a $\ar$-proof of its last formula (by the way, what does that formula say?).  $\mbox{\em Even}(x)$ should be understood as an abbreviation of  $\cle z(x=z+z)$,  and $\mbox{\em Odd}$ as an abbreviation of $\cla z(x\not=z+z)$. \vspace{7pt}

1. $\mbox{\em Even}(0)$  \ \ {\bf PA}\vspace{3pt}

2. $\mbox{\em Even}(0)\add \mbox{\em Odd}(0)$  \ \ { LC: 1}\vspace{3pt}

3.1. \ \ $\mbox{\em Even}(x)\add \mbox{\em Odd}(x)$ \ \ { Hypothesis}\vspace{3pt}

3.2. \ \ $\mbox{\em Even}(x) \mli \mbox{\em Odd}(x')$ \ \ {\bf PA}\vspace{3pt}

3.3. \ \ $\mbox{\em Odd}(x) \mli \mbox{\em Even}(x')$ \ \ {\bf PA}\vspace{3pt}

3.4. \ \ $\mbox{\em Even}(x')\add \mbox{\em Odd}(x')$ \ \ { LC: 3.1, 3,2, 3.3}\vspace{3pt}

4. $\ada x(\mbox{\em Even}(x)\add \mbox{\em Odd}(x))$ \ \ { CI: 2, 3}\vspace{3pt}

5. $\cla x\cla y\Bigl(\bigl(\mbox{\em Even}(x)\mlc \mbox{\em Even}(y)\bigr)\mld\bigl(\mbox{\em Odd}(x)\mlc \mbox{\em Odd}(y)\bigr)\mli \mbox{\em Even}(x+y)\Bigr) $ \ \ {\bf PA}\vspace{3pt}

6. $\cla x\cla y\Bigl(\bigl(\mbox{\em Even}(x)\mlc \mbox{\em Odd}(y)\bigr)\mld\bigl(\mbox{\em Odd}(x)\mlc \mbox{\em Even}(y)\bigr)\mli \mbox{\em Odd}(x+y)\Bigr) $ \ \ {\bf PA}\vspace{3pt}

7. $\cla y\bigl(\mbox{\em Even}(y)\add \mbox{\em Odd}(y) \mli \ada x(\mbox{\em Even}(x+y)\add \mbox{\em Odd}(x+y)\bigr)$  \ \ { LC: 4, 5, 6}
\end{example}

We could, of course, directly find an algorithmic solution for the problem expressed by formula 7 of the above example. But had our ad hoc methods failed to succeed --- which would likely be the case if the problem was more complex than it is --- an algorithmic solution for it or any other $\ar$-provable problem could be effectively found from a $\ar$-proof of it, as implied by the following, already mentioned theorem, establishing the status of $\ar$ as a systematic problem-solving tool: 

\begin{theorem}\label{may11} 
For any formula or sequent $X$, if $\ar\vdash X$, then  $X$ (under the standard interpretation) is computable. 

Furthermore, there is an effective procedure that takes an arbitrary extended $\ar$-proof of an arbitrary formula or sequent $X$ and constructs a BMEPM (or EPM, or HPM if you like) $\cal M$ with ${\cal M}\models X$. 
\end{theorem}

\begin{proof} We prove this theorem by induction on the lengths of deductions. As provability of formulas is a special (or easier) case of provability of sequents, considering only sequents here would be sufficient. 

Let $X=\vec{E}\fintimpl F$, and consider any $\ar$-deduction of $F$ from hypotheses $\vec{E}$. 

If $F$ is an axiom, obviously it is computable. Specifically, all axioms except Axiom 7 are true elementary formulas, and they are ``computed'' by a machine that makes no moves at all. And Axiom 7 is computed by a machine that waits for the adversary to make a move $n$, then increments $n$ by one and makes (the decimal string representing) the resulting number as its only move in the game. According to our conventions, semantically $F$ is the same as $\fintimpl F$. So, we know how to compute $\fintimpl F$. Then, by Proposition \ref{may4c}, we also know how to compute $\vec{E}\fintimpl F$.

Next, suppose $F$ is obtained from premises $G_1,\ldots,G_k$ by LC. Observe that, for each $G_i$, we have a shorter (shorter than that of $F$ from $\vec{E}$) deduction of $G_i$ from $\vec{E}$ and, of course, such a deduction can be effectively obtained from the deduction of $F$. Thus, by the induction hypothesis, we know how to compute $\vec{E}\fintimpl G_i$. Since $F$ is derived by LC, we have $\cltw\vdash G_1,\ldots,G_k\fintimpl F$ and hence, by Theorem \ref{main}, we also know how to compute 
$G_1,\ldots,G_k\fintimpl F$.\footnote{This is the only place where we rely on the condition that the proof under consideration is an extended one.  One can get by without ``extended'' is one is willing to let the algorithm search for the needed $\cltw$-proofs. Such an algorithm, however, would be dramatically more complex than the present one.} Applying Theorem \ref{april22} $k$ times (in combination with Propositions \ref{may4b} and \ref{may4d}), we then also know how to compute 
$\vec{E}\fintimpl F$.

Finally, suppose $F$ is obtained from premises $G(0)$ and a subdeduction of $G(x')$ from the hypothesis $G(x)$ by CI, so that $F$ is $\ada xG(x)$. As in the previous case, by the induction hypothesis, we know how to compute $\vec{E}\fintimpl G(0)$. Further, note that $G(x')$ has a shorter (than that of $F$ from $\vec{E}$) deduction from the hypotheses $\vec{E},G(x)$. So, by the induction hypothesis, we also know how to compute $\vec{E},G(x)\fintimpl G(x')$ and hence (by Proposition \ref{may4e}) $\vec{E},G(0)\fintimpl G(0')$. The latter, as a game, is the same as 
$\vec{E},G(0)\fintimpl G(1)$. Then, by Theorem \ref{april22} (in combination with Propositions \ref{may4b} and \ref{may4d}), we know how to compute $\vec{E}\fintimpl G(1)$. Denote by ${\cal M}_1$ the machine that computes $\vec{E}\fintimpl G(1)$. Continuing in the same way, we can construct machines ${\cal M}_2$, ${\cal M}_3$, \ldots that compute $\vec{E}\fintimpl G(2)$, $\vec{E}\fintimpl G(3)$, \ldots. Thus, for any $n$, we can effectively construct a machine ${\cal M}_n$ with 
${\cal M}_n\models \vec{E}\fintimpl G(n)$. Now, the goal problem $\vec{E}\fintimpl \ada xG(x)$ is computed by a machine that waits till the environment brings the game down to $\vec{E}\fintimpl G(n)$ for some particular $n$, after which it constructs ${\cal M}_n$ and, by simulating it (or ``turning itself into ${\cal M}_n$''), plays the rest of the game as ${\cal M}_n$ would. 
\end{proof}

\begin{exercise}\label{j24c} Show that $\ar\vdash\ada x\ada y(x=y\add x\not=y)$. 
\end{exercise}

We want to close this section by observing one fact.\footnote{Pointed out by the referee.}  In view of Axiom 7, Example  \ref{apr5b}, Example \ref{j24b} and Exercise \ref{j24c}, any interpretation for which the axioms and rules of $\ar$ are sound (and which is therefore a model of {\bf PA}) must interpret 
$'$, $+$ and $\times$ as recursive functions and $=$ as a recursive relation.  So, (one of the versions of)  Tennenbaum's well-known theorem applies and says   that the interpretation must be elementarily equivalent to the standard model. Thus, computability logic goes well beyond traditional first-order logic, not only by its computation-based semantics but also by being able to characterize the standard model of arithmetic.   

\section{Some variations of $\ar$}\label{ss12}
Among the main purposes of the present article was just to {\em introduce} CL-based applied theories, in the particular form of the CL-based arithmetic $\ar$. There is a tremendous amount of interesting and promising work in the direction of exploring such theories and their metatheories, left as a challenge for the future. In this line of research, some other versions of CL-based arithmetic might as well be at least equally interesting.

One such version is the extension of $\ar$ that we call $\artwo$. The language of the latter is no longer limited to $0,=,',+,\times$, but rather is the full language of $\cltw$, having all constants and an infinite supply of fresh predicate and function symbols for each arity, with the above  symbols having a special status. Other than that, the axioms and rules of $\artwo$ are virtually the same as those of $\ar$ (only, Axiom 8 and the two rules no longer limited to $\ar$-formulas). By {\bf a standard interpretation} we now mean any interpretation that agrees with the standard interpretation $^\dagger$ of Section \ref{ss11} on $=,',+,\times$. And let us say that a formula or sequent $X$ of the language of $\artwo$ is {\bf $\artwo$-valid} iff there is a BMEPM $\cal M$ such that, for any standard interpretation $^*$, ${\cal M}\models X^*$.  

Going back to our proof of the soundness of $\ar$, one can see that the proof of Theorem \ref{may11} never really relied on the fact that the language of $\ar$ was limited to $=,',+,\times$. So, that theorem can be  without any additional efforts strengthened to the following one: 

\begin{theorem}\label{may111} 
For any formula or sequent $X$, if $\artwo\vdash X$, then $X$ is $\artwo$-valid.  

Furthermore, there is an effective procedure that takes an arbitrary extended $\artwo$-proof of an arbitrary formula or sequent $X$ and constructs a BMEPM (or EPM, or HPM if you like) $\cal M$ such that, for every standard interpretation $^*$,  ${\cal M}\models X^*$. 
\end{theorem}

Among the reasons for wanting to consider $\artwo$ instead of $\ar$ can be the greater flexibility and convenience that it offers. Mathematicians usually do not confine themselves to strictly fixed collections of symbols such as $=,',+,\times$, and feel free to introduce new function or predicate symbols in their activities. $\artwo$ allows us to directly account for that kind of practice. Specifically, one can introduce a number of predicate and/or function symbols through definitions $E_1,\ldots,E_n$, and then prove a formula $F$ containing such symbols through proving the sequent $E_1,\ldots,E_n\fintimpl F$ in $\artwo$. 
  
As an example, below we show how all primitive-recursive functions can be proven computable in $\artwo$, with $\artwo$ thus providing a tool for actually computing any such function. Obviously the same can be done  within $\ar$ as well, as primitive-recursive functions can eventually be expressed only using $=,',+$ and $\times$. But this would be a very indirect way, with the resulting proofs leading to terrible algorithms. In contrast, $\artwo$ produces natural proofs, yilding algorithms for computing primitive-recursive functions by directly using their primitive-recursive constructions. 

We first reproduce some definitions from \cite{Kle52}.
Let $f$ be a function symbol of the indicated (by the number of explicitly shown arguments) arity. 

An {\bf absolute primitive-recursive definition} of $f$ is a formula of one of the following forms:

\begin{quote}
\begin{description}
\item[(I)] $\cla x \bigl(f(x)=x'\bigr)$.
\item[(II)] $\cla x_1\ldots\cla x_n \bigl(f(x_1,\ldots,x_n)=0\bigr)$.
\item[(III)] $\cla x_1\ldots\cla x_n \bigl(f(x_1,\ldots,x_n)=x_i\bigr)$ (some $i\in\{1,\ldots,n\}$).
\end{description}
\end{quote}

And a {\bf relative primitive-recursive definition} of $f$ is a formula of one of the following forms:

\begin{quote}
\begin{description}
\item[(IV)] $\cla x_1\ldots\cla x_n \Bigl(f(x_1,\ldots,x_n)=g\bigl(h_1(x_1,\ldots,x_n),\ldots,h_m(x_1,\ldots,x_n)\bigr)\Bigr)$.
\item[(V)] $\begin{array}{l}
\cla x_2\ldots\cla x_n \bigl(f(0,x_2,\ldots,x_n)=g(x_2,\ldots,x_n)\bigr)\ \mlc
\\
\cla x_1 \cla x_2\ldots\cla x_n \Bigl(f(x'_1,x_2,\ldots,x_n)=h\bigl(x_1,f(x_1,x_2,\ldots,x_n),x_2,\ldots,x_n\bigr)\Bigr).
\end{array}$
\end{description}
\end{quote}

We say that (IV) defines $f$ {\bf in terms of} $g,h_1,\ldots,h_m$. Similarly, we say that (V) defines $f$ in terms of $g$ and $h$. 

A {\bf primitive-recursive construction} of $f$ is a sequence $E_1,\ldots,E_k$ of formulas, where each $E_i$ is a primitive-recursive definition of some $g_i$, all such $g_i$ are distinct, $g_k=f$ and, for each $i$, $E_i$ is either an absolute primitive-recursive definition of $g_i$, or a relative primitive-recursive definition of $g_i$ in terms of some $g_j$s with $j<i$.  

\begin{proposition}\label{may12}
Let $f$ be an $n$-ary function symbol, and $\vec{E}=E_1,\ldots,E_k$ a primitive-recursive construction of $f$. Then $\artwo\vdash \vec{E}\fintimpl\ada x_1\ldots\ada x_n\ade y\bigl(f(x_1,\ldots,x_n)=y\bigr)$. 
\end{proposition}

\begin{proof} Assume the conditions of the proposition. We prove it by induction on $k$, considering five possibilities for $E_k$. As hypotheses of the following deductions, we only include those $E_i$ ($i<k$) that are explicitly used. The other $E_j$, of course, can be vacuously added to the list of hypotheses.\vspace{7pt}

{\em Case (I)}: $E_k$ has the form (I).  Below is a $\artwo$-proof of the desired sequent:\vspace{5pt}

\noindent 1. $\cla x \bigl(f(x)=x'\bigr)$  \ \ { Hypothesis}\vspace{3pt}

\noindent 2. $\ada x\ade y(y=x')$ \ \ { Axiom 7}\vspace{3pt}

\noindent 3. $\ada x\ade y\bigl(f(x)=y\bigr)$ \ \ { LC: 1, 2}\vspace{10pt}

{\em Case (II)}: $E_k$ has the form (II).  Below is a $\artwo$-proof of the desired sequent.\vspace{5pt}

\noindent 1. $\cla x_1\ldots\cla x_n \bigl(f(x_1,\ldots,x_n)=0\bigr)$  \ \ { Hypothesis}\vspace{3pt}

\noindent 2. $\ada x_1\ldots\ada x_n\ade y \bigl(f(x_1,\ldots,x_n)=y\bigr)$ \ \ { LC: 1}\vspace{10pt}

{\em Case (III)}: $E_k$ has the form (III).  Below is a $\artwo$-proof of the desired sequent.\vspace{5pt}

\noindent 1. $\cla x_1\ldots\cla x_n \bigl(f(x_1,\ldots,x_n)=x_i\bigr)$ \ \ { Hypothesis}\vspace{3pt}

\noindent 2. $\ada x_1\ldots\ada x_n \ade y\bigl(f(x_1,\ldots,x_n)=y\bigr)$ \ \ { LC: 1}\vspace{10pt}

{\em Case (IV)}: $E_k$ has the form (IV).  Below is a (lazy) $\artwo$-proof of the desired sequent.\vspace{5pt}

\noindent 1. \ \ \ \ \ $\cla x_1\ldots\cla x_n \Bigl(f(x_1,\ldots,x_n)=g\bigl(h_1(x_1,\ldots,x_n),\ldots,h_m(x_1,\ldots,x_n)\bigr)\Bigr)$ \ \ { Hypothesis}\vspace{3pt}

\noindent 2. \ \ \ \ \ $\ada x_1\ldots\ada x_m\ade y \bigl(g(x_1,\ldots,x_m)=y\bigr)$ \ \ { Provable by the induction hypothesis}\vspace{3pt}

\noindent 3. \ \ \ \ \ $\ada x_1\ldots\ada x_n\ade y \bigl(h_1(x_1,\ldots,x_n)=y\bigr)$ \ \ { Provable by the induction hypothesis}\vspace{3pt}

\noindent{\LARGE $\cdots$}

\noindent m+2. $\ada x_1\ldots\ada x_n\ade y \bigl(h_m(x_1,\ldots,x_n)=y\bigr)$ \ \ { Provable by the induction hypothesis}\vspace{3pt}

\noindent m+3.  $\ada x_1\ldots\ada x_n\ade y \bigl(f(x_1,\ldots,x_n)=y\bigr)$ \ \ { LC: 1, \ldots, m+2}\vspace{5pt}

{\em Case (V)}: $E_k$ has the form (V).  Below is a (lazy) $\artwo$-proof of the desired sequent.\vspace{10pt}

\noindent 1. $\begin{array}{l}
\cla x_2\ldots\cla x_n \bigl(f(0,x_2,\ldots,x_n)=g(x_2,\ldots,x_n)\bigr)\ \mlc
\\
\cla x_1 \cla x_2\ldots\cla x_n \Bigl(f(x'_1,x_2,\ldots,x_n)=h\bigl(x_1,f(x_1,x_2,\ldots,x_n),x_2,\ldots,x_n\bigr)\Bigr)
\end{array}$ \ \ { Hypothesis}\vspace{3pt}

\noindent 2. \ $\ada x_2\ldots\ada x_n\ade y \bigl(g(x_2,\ldots,x_n)=y\bigr)$ \ \ { Provable by the induction hypothesis}\vspace{3pt}

\noindent 3. \ $\ada x_1\ada z\ada x_2\ldots\ada x_n\ade y \bigl(h(x_1,z,x_2,\ldots,x_n)=y\bigr)$ \ \ { Provable by the induction hypothesis}\vspace{3pt}

\noindent 4. \ $\ada x_2\ldots\ada x_n\ade y \bigl(f(0,x_2,\ldots,x_n)=y\bigr)$ \ \ { LC: 1, 2}\vspace{3pt}

\noindent 5.1. \ \ \ \ $\ada x_2\ldots\ada x_n\ade y \bigl(f(x_1,x_2,\ldots,x_n)=y\bigr)$ \ \ { Hypothesis}\vspace{3pt}

\noindent 5.2. \ \ \ \ $\ada x_2\ldots\ada x_n\ade y \bigl(f(x'_1,x_2,\ldots,x_n)=y\bigr)$ \ \ { LC: 1, 3, 5.1}\vspace{3pt}

\noindent 6. \ $\ada x_1\ada x_2\ldots\ada x_n\ade y \bigl(f(x_1,x_2,\ldots,x_n)=y\bigr)$ \ \ { CI: 4, 5}\vspace{5pt}

\end{proof}

The other CL-based system of arithmetic that we want to introduce before closing this section is $\arthree$. While $\artwo$ is more expressive than $\ar$, $\arthree$ is a modification of $\ar$ in the opposite direction: the language of $\arthree$ is obtained from the language of $\ar$ by forbidding the blind quantifiers $\cla$ and $\cle$. The rules of inference of $\arthree$ are the same LC and CI as in $\ar$, only now restricted to $\cla,\cle$-free formulas. 
And the axioms are the following:\vspace{-3pt}

\begin{enumerate}
\item $\ada x(0\not=x')$\vspace{-3pt}
\item $\ada x\ada y(x'=y'\mli x=y)$\vspace{-3pt}
\item $\ada x(x+0=x)$\vspace{-3pt}
\item $\ada x\ada y\bigl(x+y'=(x+y)'\bigr)$\vspace{-3pt}
\item $\ada x(x\times 0=0)$\vspace{-3pt}
\item $\ada x\ada y\bigl(x\times y'=(x\times y)+x\bigr)$\vspace{-3pt}
\item $\ada x\ade y(y=x')$\vspace{-3pt}
\end{enumerate}
 
In view of the obvious fact that $\cltw$ always proves $\cla xE(x)\fintimpl \ada x E(x)$, all of the axioms of $\arthree$, which differ from the corresponding axioms of $\ar$ only in that they are $\ada$- rather than $\cla$-prefixed, are provable in $\ar$. This, in turn, implies that $\arthree$, just like $\artwo$, inherits the soundness of $\ar$ in the strong form of Theorem \ref{may11}.

From the philosophical point of view, the potential interest in $\arthree$ is related to the fact that it is a perfectly constructive system of arithmetic and, unlike some other systems with constructivistic claims such as Heyting's intuitionistic-calculus-based arithmetic, has a clear and convincing constructive semantics. What makes $\arthree$ ``perfectly constructive'' is that it gets rid of the only potentially ``dubious'' operators $\cla$ and $\cle$, with all the remaining operators being fully immune to any doubts or criticism from even the most radical constructivistic point of view.   A relevant observation to be made here is that forbidding $\cla,\cle$ in the formulas of $\arthree$ automatically propagates to the underlying logic $\cltw$ as well: it is not hard to see that, when the premises and conclusion of LC are $\cla,\cle$-free, then so are all intermediate steps performed within $\cltw$ when justifying an application of LC. That means that, in fact, the underlying logic of $\arthree$ is the $\cla,\cle$-free fragment of $\cltw$ rather than the full $\cltw$.

On the mathematical side, an important feature of $\arthree$ is that its concept of ``truth'' (=computability), unlike that of $\ar$ or just {\bf PA}, can be expressed in the language of {\bf PA}. Specifically, one could show that the arithmetical complexity of the predicate of ``truth'' for $\arthree$-formulas is $\Sigma_3$. Replacing the classical quantifiers in such a predicate by their choice counterparts $\ada$ and $\ade$ yields a formula which, in a sense, expresses the ``truth'' predicate of $\arthree$ in the language of $\arthree$ itself. This unusual phenomenon may yield unusual effects in the metatheory of $\arthree$, which can make $\arthree$ an interesting subject for metainvestigations.  
 
\section{The closure of static games under dfb-reduction}\label{ss13}
This section should be treated as an appendix, which many readers may (safely) decide to skip.  In it we 
prove a technical but important fact, according to which $\fintimpl$ preserves the static property of games. Not passing the test for preserving the static property would disqualify any game operation from being considered within the current framework of CL, as only static games are ``well behaved'', and many steps in our earlier proofs (specifically, the proofs of the closure properties of Section \ref{ss7}, the proof of the soundness and completeness of $\cltw$, or the proofs of the soundness of $\ar$, $\artwo$, $\arthree$) would fail if the underlying games were not static. 

Let us remember the key relevant definitions first. We say that a run $\Delta$ is a {\bf $\pp$-delay} of a run $\Gamma$ iff the following two conditions are satisfied:
\begin{itemize} 
\item For either player $\xx\in\{\twg,\tlg\}$, erasing all $\xx$-labeled moves in $\Gamma$ results in the same run as erasing all $\xx$-labeled moves in $\Delta$.
\item For all $k$ and $n$, if the $k$th $\pp$-labeled move is made later than (is to the right of) the $n$th $\oo$-labeled move in $\Gamma$, then so is it in $\Delta$.
\end{itemize} 

{\bf $\oo$-delay} is defined similarly, with $\pp$ and $\oo$ interchanged. 

Now, we say that a constant game $A$ is {\bf static} iff, for any player $\xx$, run $\Gamma$ and $\xx$-delay $\Delta$ of $\Gamma$, the following two conditions are satisfied:
\begin{enumerate}
\item If $\Gamma$ is not a $\xx$-illegal run of $A$, then neither is $\Delta$.
\item If $\Gamma$ is a $\xx$-won run of $A$, then so is $\Delta$.
\end{enumerate}

And a nonconstant game is considered static iff so are all of its instances.

Remark: Another equivalent approach, taken in all earlier papers on CL except \cite{Japic}, only stipulates the second condition in the definition of static games. That approach, however, assumes existence of a move $\spadesuit$ that is illegal in every position of every game. As it turns out (Lemma 4.7 of \cite{Jap03}), the first condition of our present definition of static games is then automatically implied by the second condition, so there is no need for explicitly stating it. Without stipulating the existence of an always-illegal move, however, both conditions are necessary.  

Our Definition \ref{op1} of $\fintimpl$ was limited to finite-depth games. The only reason for adopting this limitation was the desire to keep things simple, as finite-depth games were sufficient for our present treatment: every formula of $\cltw$ expresses a finite-depth game,  as elementary games are finite-depth, and the operations $\gneg,\mlc,\mld,\adc,\add,\ada,\ade,\cla,\cle$ (but not $\fintimpl$) preserve the finite-depth property. However, for possible future needs, in this appendix-style technical section where we no longer care about simplicity, it would not hurt to generalize $\fintimpl$ to all games  before proving that it preserves the static property of games. The following definition does this job.

Here are some terminological and notational conventions for Definition \ref{ffff}. 
The {\bf predecessor} of a nonempty finite bit string $w$ is the longest proper prefix of $w$. For a  bit string $w$ and run $\Gamma$, $\Gamma^{\preceq w}$ means the result of deleting from $\Gamma$ all labmoves except those that look like $\xx u.\alpha$ for some $u\preceq w$, and then further deleting ``$u.$'' in each such $\xx u.\alpha$. Similarly,
$\Gamma^{S.}$ means the result of deleting from $\Gamma$ all labmoves except those that look like $\xx S.\alpha$, and then further deleting ``$S.$'' in each such $\xx S.\alpha$.

\begin{definition}\label{ffff}

Let  $A_1,\ldots,A_n$ and $B$  be  constant games, and $\cal T$ a tree of games with yield $\seq{A_1,\ldots,A_n}$ ($n\geq 1$).  Let $w_1,\ldots,w_n$ be the addresses of $A_1,\ldots,A_n$ in $\cal T$, respectively. The game   ${\cal T}\fintimpl B$ is defined by:
\begin{description}
\item[$\legal{}{}$:] $\Gamma$ is a legal run of ${\cal T}\fintimpl B$ iff the following conditions are satisfied:
\begin{enumerate}
\item Every labmove of $\Gamma$ looks like $\xx S.\beta$, $\pp \col{w}$ or $\xx w.\beta$, where $\xx$ is either player, $\beta$ is a move, and $w$ is a finite (possibly empty) bit string. 
\item Whenever $\Gamma$ contains $\pp\col{w}$, it does not have any other occurrences of the same labmove, and either $w$ is one of $w_1,\ldots,w_n$, or the labmove $\pp\col{w}$ is preceded by $\pp\col{u}$,\footnote{Here and below, ``preceded'' does not mean ``immediately preceded''. Rather, it means that $\Gamma$ looks like $\seq{\ldots\pp\col{u}\ldots\pp\col{w}\ldots}$.} where $u$ is the predecessor of $w$. We call labmoves of the form $\pp\col{w}$ {\bf replicative}.  
\item  Whenever $\Gamma$ contains $\xx w.\beta$,  either $w$ is a (not necessarily proper) prefix of one of $w_1,\ldots,w_n$, or else the labmove $\xx w.\beta$ is preceded by $\pp\col{u}$, where $u$ is the predecessor of $w$. 
\item $\Gamma^{S.}$ is a legal run of $B$.
\item For any $i\in\{1,\ldots,n\}$ and any infinite bit string $u$, $\Gamma^{\preceq w_iu}$ is a legal run of $\gneg A_i$.
\end{enumerate}

\item[$\win{}{}$:] A legal run $\Gamma$ of ${\cal T}\fintimpl B$ is won by $\pp$ iff $\Gamma$ contains only finitely many replicative labmoves, and either $\win{B}{}\seq{\Gamma^{S.}}=\pp$ or, for some $i\in\{1,\ldots,n\}$ and some infinite bit string $u$,\footnote{In fact, considering only certain ``sufficiently long'' finite bit strings would do the job here, but why bother.} 
$\win{\gneg A_i}{}\seq{\Gamma^{\preceq w_iu}}=\pp$.

\end{description}
\end{definition}

It is not hard to see that, when restricted to finite-depth games, the above definition of $\fintimpl$ is equivalent to Definition \ref{op1}.

\begin{proposition}\label{static}
Let $A_1,\ldots,A_n$ and $B$ be arbitrary  static games, and $\cal T$ be a tree of games with yield $\seq{A_1,\ldots,A_n}$ ($n\geq 1$). Then the game ${\cal T}\fintimpl B$ is also static. 
\end{proposition}

\begin{proof} Suppose $A_1,\ldots,A_n$, $B$ and $\cal T$ are as in the assumption of the lemma. We may also safely assume that the $n+1$ games are constant, or else pick an arbitrary valuation $e$ and replace them with $e[A_1],\ldots,e[A_n],e[B]$. Let $\xx$ be either player, $\Gamma$ any run, and $\Delta$ any $\xx$-delay of $\Gamma$.\vspace{5pt}

{\em Verifying the first condition of the definition of static games}. 
Assume $\Delta$ is a $\xx$-illegal run of ${\cal T}\fintimpl B$. We want to show that then so is $\Gamma$. Below, when we simply say ``legal'', ``$\xx$-illegal'', etc., we mean (being a) legal, $\xx$-illegal, etc. run of ${\cal T}\fintimpl B$.

Let $\seq{\Psi,\xx\alpha}$ be the shortest $\xx$-illegal initial segment of $\Delta$. Let $\seq{\Phi,\xx\alpha}$ be the shortest initial segment of $\Gamma$ containing all the $\xx$-labeled moves of $\seq{\Psi,\xx\alpha}$, and let $\Theta$ be the sequence of those $\overline{\xx}$-labeled moves of $\Psi$ that are not in $\Phi$. We obviously have 
\begin{equation}\label{foo61}
\mbox{\em $\seq{\Psi,\xx\alpha}$ is a $\xx$-delay of $\seq{\Phi,\xx\alpha,\Theta}$.}
\end{equation}

If $\Phi$ is  $\xx$-illegal, then so is $\Gamma$ and we are done. Assume now that 

\begin{equation}\label{m11}
\mbox{\em $\Phi$ is not $\xx$-illegal.}
\end{equation}

We claim that then

\begin{equation}\label{fooo}
\mbox{\em $\Phi$ is legal.}
\end{equation}
Indeed, suppose that this was not the case. Then, by (\ref{m11}), $\Phi$ should be $\overline{\xx}$-illegal. This would make $\Gamma$ a $\overline{\xx}$-illegal run  with $\Phi$ as an illegal initial segment which is shorter than $\seq{\Psi,\xx\alpha}$. Then, by the induction hypothesis, any run for which $\Gamma$ is a $\overline{\xx}$-delay, would be $\overline{\xx}$-illegal. But the fact that $\Delta$ is a $\xx$-delay of $\Gamma$ obviously implies that $\Gamma$ is a $\overline{\xx}$-delay of $\Delta$ (Lemma 4.6 of \cite{Jap03}). So, $\Delta$ would be $\overline{\xx}$-illegal, which is a contradiction because, according to our assumptions, $\Delta$ is $\xx$-illegal. 

We are continuing our proof. There are five possible reasons to why $\seq{\Psi,\xx\alpha}$ is  $\xx$-illegal (while $\Psi$ being legal):

{\em Reason 1}: $\xx\alpha$ does not have the form $\xx S.\beta$, $\pp\col{w}$ or $\xx w.\beta$. But then, for the same reason, $\seq{\Phi,\xx\alpha}$ is illegal. This, in view of (\ref{fooo}), means that $\seq{\Phi,\xx\alpha}$ is $\xx$-illegal. Hence $\Gamma$, as an extension of   $\seq{\Phi,\xx\alpha}$, is also $\xx$-illegal, as desired. 

{\em Reason 2}: $\xx\alpha$ is $\pp\col{w}$, but it violates condition 2 of the {\bf Lr} clause of Definition \ref{ffff}. It is not hard to see that then, for the same reason, $\seq{\Phi,\xx\alpha}$ is illegal.  This, in turn, as in the previous case, implies that $\Gamma$ is $\xx$-illegal. 
  
{\em Reason 3}: $\xx\alpha$ is $\pp w.\beta$, but it violates condition 3 of the {\bf Lr} clause of Definition \ref{ffff}. Again, for this very reason, $\seq{\Phi,\xx\alpha}$ can be seen to be  illegal.  This, as in the previous cases, implies that $\Gamma$ is $\xx$-illegal.

{\em Reason 4}: $\seq{\Psi,\xx\alpha}^{S.}$ is an illegal run of $B$. This means it is a $\xx$-illegal run of $B$, because we know that $\Psi$ is a legal run of 
${\cal T}\fintimpl B$. Obviously (\ref{foo61}) implies that $\seq{\Psi,\xx\alpha}^{S.}$ is a $\xx$-delay of $\seq{\Phi,\xx\alpha,\Theta}^{S.}$. Hence, as $B$ is static, $\seq{\Phi,\xx\alpha,\Theta}^{S.}$ is also a $\xx$-illegal run of $B$. From here, taking into account that $\Theta$ does not have any $\xx$-labeled moves, we find that $\seq{\Phi,\xx\alpha}^{S.}$ is a $\xx$-illegal run of $B$. This, together with (\ref{fooo}), obviously implies that $\seq{\Phi,\xx\alpha}$ is a $\xx$-illegal run of ${\cal T}\fintimpl B$, and hence so is its extension $\Gamma$, as desired.

{\em Reason 5}: For some $i\in\{1,\ldots,n\}$ and some infinite bit string $u$, $\seq{\Psi,\xx\alpha}^{\preceq w_iu}$ is not a legal run of $\gneg A_i$. Fix these $i$ and $u$. This case is very similar to the previous one. Since $\Psi$ is a legal run of 
${\cal T}\fintimpl B$, we should have that $\seq{\Psi,\xx\alpha}^{\preceq w_iu}$ is a $\xx$-illegal run of $\gneg A_i$. Also, (\ref{foo61}) implies that $\seq{\Psi,\xx\alpha}^{\preceq w_iu}$ is a $\xx$-delay of $\seq{\Phi,\xx\alpha,\Theta}^{\preceq w_iu}$. Therefore, as $A_i$ and hence $\gneg A_i$ are static, $\seq{\Phi,\xx\alpha,\Theta}^{\preceq w_iu}$ is a $\xx$-illegal run of it. Further reasoning as in the previous case,  we find that  $\seq{\Phi,\xx\alpha}$  and hence $\Gamma$ is a $\xx$-illegal run of ${\cal T}\fintimpl B$.
\vspace{7pt}

{\em Verifying the second condition of the definition of static games}. Here, again, when we simply say ``legal'', ``$\xx$-illegal'', ``$\xx$-won'', etc., we mean (being a) legal, $\xx$-illegal, $\xx$-won etc. run of ${\cal T}\fintimpl B$.

Assume $\Gamma$ is won by $\xx$. We want to show that then so is $\Delta$. 

If $\Delta$ is $\overline{\xx}$-illegal, then it is won by $\xx$ and we are done. So, assume that 
$\Delta$ is not $\overline{\xx}$-illegal. But, as noted earlier, $\Gamma$ is a $\overline{\xx}$-delay of $\Delta$ and, therefore, in view of  the already proven fact that the first condition of the definition of static games is satisfied, we find that $\Gamma$ is not $\overline{\xx}$-illegal, either.
$\Gamma$ also cannot be $\xx$-illegal, for otherwise it would not be won by $\xx$. Consequently, $\Delta$ cannot be $\xx$-illegal either, for otherwise $\Gamma$ would be $\xx$-illegal. Thus, we have narrowed down our considerations to the case when both $\Gamma$ and $\Delta$ are legal.

If the reason of $\Gamma$'s being won by $\xx$ is that $\Gamma$ has infinitely many replicative labmoves and $\xx=\oo$, then $\Delta$ is won by $\xx$ for exactly the same reason.

Assume now $\Gamma$ only has a finite number of replicative labmoves (and hence so does $\Delta$). 

Suppose $\Gamma^{S.}$ is  a $\pp$-won run of $B$, so that $\xx$ (the winner in $\Gamma$) is $\pp$. $\Delta$'s being a $\pp$-delay of $\Gamma$ implies that $\Delta^{S.}$ is a $\pp$-delay of $\Gamma^{S.}$, and hence, as the latter is a $\pp$-won run of the static game $B$, so is the former. This, in turn, implies that $\Delta$ is a $\pp$-won (i.e., $\xx$-won) run of   ${\cal T}\fintimpl B$, as desired.

Suppose now $\Gamma^{S.}$ is not a $\pp$-won run of $B$. Then, our assumption that $\Gamma$ is a $\xx$-won run of  ${\cal T}\fintimpl B$ can be seen to imply that, for each $i\in\{1,\ldots,n\}$ and each infinite bit string $u$  
(if $\xx=\oo$), or some such $i$ and  $u$ (if $\xx=\pp$),  $\Gamma^{\preceq w_iu}$ is a $\xx$-won run of $\gneg A_i$.  Taking into account that $\Delta^{\preceq w_iu}$ is obviously a $\xx$-delay of $\Gamma^{\preceq w_iu}$ and that $\gneg A_i$ is static, the above, in turn, implies that for each $i$ and  $u$ (if $\xx=\oo$), or some $i$ and $u$ (if $\xx=\pp$), $\Delta^{\preceq w_iu}$ is a $\xx$-won run of $\gneg A_i$. In addition, if $\xx=\oo$, then $\Delta^{S.}$ is a $\xx$-won run of $B$, because so is $\Gamma^{S.}$, $B$ is static and $\Delta^{S.}$ is a $\xx$-delay of $\Gamma^{S.}$. All this implies the desired conclusion that $\Delta$ is a $\xx$-won run of ${\cal T}\fintimpl B$. 
\end{proof}


\begin{thebibliography}{99}


\bibitem{Jap03} G. Japaridze. {\em Introduction to computability logic}. {\bf Annals of Pure and Applied Logic} 123 (2003), No.1-3, pp. 1-99.

\bibitem{Japtocl1} G. Japaridze. {\em Propositional computability logic I}. {\bf ACM Transactions on Computational Logic} 7 (2006), No.2, pp. 302-330.

\bibitem{Japtocl2} G. Japaridze. {\em Propositional computability logic II}. {\bf ACM Transactions on Computational Logic} 7 (2006), No.2, 
pp.  331-362.

\bibitem{Japtcs} G. Japaridze. {\em From truth to computability I}. {\bf Theoretical Computer Science} 357 (2006), No.1-3, pp. 100-135.

\bibitem{Cirq} G. Japaridze. {\em Introduction to cirquent calculus and abstract resource semantics}. {\bf Journal of Logic and Computation} 16 (2006), No.4, pp. 489-532.



\bibitem{Japic} G. Japaridze. {\em Computability logic: a formal theory of interaction}. In: {\bf Interactive Computation: The New Paradigm}. D. Goldin, S. Smolka and P. Wegner, eds. Springer  2006, pp. 183-223. 
   

\bibitem{Japjsl} G. Japaridze. {\em The logic of interactive Turing reduction}. {\bf Journal of Symbolic Logic} 72 (2007), No.1, pp. 243-276. 


\bibitem{Japtcs2} G. Japaridze. {\em From truth to computability II}. {\bf Theoretical Computer Science} 379 (2007), No.1-2, pp. 20-52.

\bibitem{int1} G. Japaridze. {\em Intuitionistic computability logic}. {\bf Acta Cybernetica} 18 (2007), No.1, pp. 77--113.  

\bibitem{Propint} G. Japaridze. {\em The intuitionistic fragment of computability logic at the propositional level}. {\bf Annals of Pure and Applied Logic} 147 (2007), No.3, pp. 187-227. 

\bibitem{Japdeep} G. Japaridze. {\em Cirquent calculus deepened}. {\bf Journal of Logic and Computation} 18 (2008), No.6, pp. 983-1028.  

\bibitem{Japseq} G. Japaridze. {\em Sequential operators in computability logic}. {\bf Information and Computation} 206 (2008), No.12, pp. 1443-1475.   

\bibitem{Japfin} G. Japaridze. {\em In the beginning was game semantics}. In: {\bf Games: Unifying Logic, Language, and Philosophy}. O. Majer,
A.-V. Pietarinen and T. Tulenheimo, eds. Springer 2009, pp. 249-350.   

\bibitem{Japfour} G. Japaridze. {\em Many concepts and two logics of algorithmic reduction}. {\bf Studia Logica} 91 (2009), No.1,  pp. 1-24. 


\bibitem{Kle52} S.C. Kleene. {\bf Introduction to Metamathematics.} D. van Nostrand Company, New York / Toronto, 1952.
\end{thebibliography}
\end{document}